\documentclass[twocolumn,superscriptaddress,floatfix,APS]{revtex4}
\usepackage{graphicx}

\begin{document}
\title{Multi-flavor bosonic Hubbard models in the first excited Bloch band of an optical lattice}
\author{A. Isacsson}
\affiliation{NORDITA, Blegdamsvej 17, Copenhagen Oe, DK-2100, Denmark}
\affiliation{Department of Physics, Yale University, P.O. Box 208120, New Haven, CT 06520-8120}
\author{S. M. Girvin}
\affiliation{Department of Physics, Yale University, P.O. Box 208120, New Haven, CT 06520-8120}

\date{\today}
\begin{abstract}
We propose that by exciting ultra cold atoms from the zeroth to the
first Bloch band in an optical lattice, novel multi-flavor bosonic
Hubbard Hamiltonians can be realized in a new way. In these systems,
each flavor hops in a separate direction and on-site exchange terms
allow pairwise conversion between different flavors. Using band
structure calculations, we determine the parameters entering these
Hamiltonians and derive the mean field ground state phase diagram
for two effective Hamiltonians (2D, two-flavors and 3D, three
flavors).
 Further, we estimate the stability of atoms in the
first band using second order perturbation theory and find
lifetimes that can be considerably \mbox{(10-100 times)} longer
than the relevant time scale associated with inter-site hopping
dynamics, suggesting that quasi-equilibrium can be achieved in
these meta-stable states.
\end{abstract}
\maketitle

\section{Introduction}
\label{sec:intro}
The possibility to trap and manipulate the atoms in a
Bose-Einstein Condensate using standing wave laser
beams~\cite{Burger:2001,Cataliotti:2001,Jaksch:1998,Kasevich:2001,Bloch:2002,Bloch2:2002}
has led to a renewal of the interest in basic solid state models.
In such systems, the atoms experience a periodic potential from an
optical lattice leading to formation of band structure in the
energy spectrum. These bands have been investigated in
experiments~\cite{Phillips:2002}.

In the spectroscopy experiments in Ref.~\cite{Phillips:2002}, the
atoms experienced a periodic potential in only one direction, being
free to move on a much larger length scale in the other directions.
This implied that interactions between atoms could be ignored. If
the atoms are confined to reside on the sites of a lattice in three
dimensions, interactions become important. As a result, it was shown
theoretically~\cite{Jaksch:1998}, and subsequently also
experimentally~\cite{Bloch:2002}, that a system of interacting cold
atoms, residing in the lowest Bloch band of the periodic potential,
maps onto a bosonic Hubbard model. This model is of great
theoretical interest since it exhibits a quantum phase
transition~\cite{Fisher:1989,Sachdev,Oosten1:2001,Oosten2:2003}
between ground states where the atoms are localized (Mott-Insulator)
and where they are delocalized (superfluid) as the strength of the
hopping relative to the inter-atomic interaction is varied.  The
dynamics of particles under the influence of changes in the
Hamiltonian (such as lattice tilts or rapid changes in the particle
interaction strength) has also proved interesting
\cite{Kasevich:2001,Bloch:2002,Bloch2:2002,sengupta,polkovnikov}.

Another development is an interest in the idea of mixing bosonic
atoms of different flavors in the
lattice~\cite{Chen:2003,Kuklov:2002,Kuklov:2003,Altman:2003,Duan:2002,Cha:2005}.
Several ways of achieving multiple flavors have been suggested
including using atoms of different species and exploiting
different internal atomic states.

So far, experiments on strongly interacting atoms in three
dimensional optical lattices have been restricted to atoms in the
lowest (zeroth) Bloch band.  Recently Scarola and DasSarma
considered the possibility of novel supersolid phases within the
first excited Bloch band of an optical lattice.
\cite{ScarolaDasSarma}

In this paper, the theory of atoms in the two lowest (zeroth and
first) Bloch bands of a three dimensional optical lattice is
considered. We show here, that due to the lack of available phase
space for the decay products, such excited states can (in some
parameter ranges) have life times much longer than the
characteristic time scales for inter-site hopping. Thus it should
be possible to establish quasi-equilibrium within the manifold of
these metastable states.

\begin{figure}[t]
{\includegraphics[width=7cm]{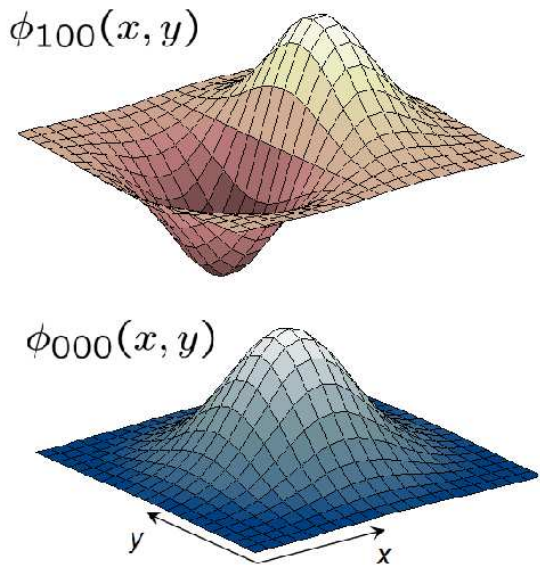}}
\caption{
(Color online) On-site Wannier wave functions in the Harmonic
oscillator approximation. The localized wave functions are to a
good approximation described by harmonic oscillator wave functions
localized in each well. Above is drawn the wave functions
$\phi_{(0,0,0)}({\mathbf r})$ (plotted in the plane $z=0$) formed
by the zeroth band Bloch functions and the wave function
$\phi_{(1,0,0)}({\mathbf r})$ formed by the zeroth band Bloch
functions in the y- and z- directions and the Bloch functions from
the first band in the x-direction.  These are approximately
harmonic oscillator states, $\phi_{(0,0,0)}({\mathbf r})\sim
\exp[-\alpha (x^2+y^2+z^2)]$, and $\phi_{(1,0,0)}({\mathbf r})\sim
x\exp[-\alpha (x^2+y^2+z^2)]$ where the parameter $\alpha$ is
determined by the curvature of the optical lattice potential near
its minima. Similarly $\phi_{(0,1,0)}({\mathbf r})\sim y \phi_{(0,0,0)}({\mathbf r})$
and $\phi_{(0,0,1)}({\mathbf r})\sim z \phi_{(0,0,0)}({\mathbf r})$.
\label{fig:wavefuns}}
\end{figure}

We find that it is possible in this way to realize novel effective
multi-species bosonic Hubbard Hamiltonians. Depending on the
choice of lattice depths the number of degenerate bands varies and
we find effective models involving $n$ flavors of bosons, where
$n$ can be 1, 2, or 3. These flavors correspond to the three
different possible nodal planes in the excited state wave function
such as the one illustrated in Fig.~(\ref{fig:wavefuns}). We will
show that a characteristic of these Hamiltonians is that (to a
good approximation) each flavor can only hop in one direction
(\emph{i.e.}, $X$ (nodal plane) particles can only hop in the $x$
direction, etc.). Neglecting interactions we would then have $n$
interpenetrating one-dimensional free bose gases, one for each
column (or row) in the lattice.  Allowing intra-species
interactions converts these one-dimensional gases into Luttinger
liquids (or, if the interactions are strong enough, and the mean
particle number per site is an integer, into Mott insulators).
We show below that, besides intra-species interactions, the full
interaction also includes on-site inter-species conversion terms
that allow atoms to change flavor in pairs. Thus for example, two
$X$ particles constrained to move along a single $x$ column can
collide, turn into $Y$ particles and move away along a $y$ column.
Such processes lead to novel quantum dynamics for this coupled set
of interpenetrating Luttinger liquids.

As will be seen, the anisotropic nature of the hopping in
conjunction with the pairwise conversion leads to Hamiltonians
with an infinite but subextensive set of $Z_2$-gauge symmetries
intermediate between local and global. Such infinite symmetries
have been found in certain frustrated spin
models~\cite{Moore1,Moore2,Nussinov1,Nussinov2} and in a `bose
metal' model~\cite{Paramekanti} and are known to cause dimensional
reduction in some cases.~\cite{Moore1,Moore2,Nussinov2} We will
see below how this dimensional reduction appears in a simple way
in this system.

A related {\em global} $Z_2$ symmetry and associated Ising order
parameter appear in problems involving boson pairing due to
attractive interactions mediated by Feshbach resonances. In that
case the symmetry appears due to a conversion term that connects
pairs of bosons with a distinct molecular field.  This can lead to
exotic states in which pairs of bosons are condensed but single
bosons are {\em not} and in which half vortices are permitted
\cite{Romans,Radzihovsky}

Further, due to strong interatomic repulsion, the ground state in 3D
(three flavors)
breaks a kind of chiral symmetry and displays an additional
accidental ground state degeneracy at the mean field level. A
similar situation occurs for special parameter values in frustrated
XY-models, where parallel zero energy domain walls can be
inserted~\cite{Korshunov}. The outline of this paper is as follows:
In section~\ref{sec:Lattice_Ham} the appropriate generalization of
the bosonic Hubbard model is introduced along with numerical values
of the parameters entering the Hamiltonians obtained from
band-structure calculations for various lattice depths. Then, in
section~\ref{sec:effective_hams}, the aforementioned effective
Hamiltonians for atoms in the first band are derived for three
particular choices of relative lattice depths in the $xyz-$
directions. Using simple mean-field theory we sketch the ground
state phase-diagrams in section~\ref{sec:MFT} and in
section~\ref{sec:interference} we discuss how the superfluid phases
are reflected in the interference pattern in an experimental
situation. Finally, in section~\ref{sec:lifetime}, treating the
interaction perturbatively to second order, we estimate the lifetime
of a population inverted state (all atoms residing entirely in the
first excited band).

\section{General Lattice Hamiltonian}
\label{sec:Lattice_Ham}
The starting point is the Hamiltonian for weakly interacting
bosons of mass $m$ in an external potential~\cite{Leggett:2001}
\begin{eqnarray}
\hat{H}&=&\int d^3{\bf x}\, \hat{\psi}^\dagger({\bf
x})\left(-\frac{\hbar^2}{2m}\nabla^2+V_O({\bf
x})+V_T({\bf x})\right)\hat{\psi}({\bf x})\nonumber \\
&+&\frac{1}{2}\frac{4\pi a_s \hbar^2}{m}\int d^3 {\bf x}\,
\hat{\psi}^\dagger({\bf x})\hat{\psi}^\dagger({\bf
x})\hat{\psi}({\bf x})\hat{\psi}({\bf x}),
\label{eq:hamiltonian}
\end{eqnarray}
where $a_s$ is the $s$-wave scattering length. The external
potential has two contributions $V_O$ and $V_T$ corresponding to
the lattice potential and the magnetic trapping potential.
Denoting the wavelength of the lasers by $\lambda\equiv 2a$, $a$
being the lattice spacing, the former can be written
$$
V_O({\bf x})=\sum_{i=x,y,z}V_{0i}\sin^2\left(\frac{2\pi}{\lambda}x_i\right),
\{x_i\}_{i=x,y,z}=(x,y,z).
$$
The positional dependence of the magnetic trapping potential
$V_T=\frac{1}{2}m\sum\limits_{i=x,y,z}\Omega_i^2 x_i^2,$ is much
weaker than that of the lattice, \emph{i.e.} $\Omega_i\ll
\frac{2\pi}{\lambda}\sqrt{\frac{2V_{0i}}{m}}$ and will be ignored
in the remainder of this paper. One should be aware though, that
this term has been shown to influence for instance the phase
diagram of the single flavor bosonic Hubbard
model~\cite{Batrouni:2002,Pupillo:2003}.

For the cubic lattices considered here, the Wannier functions
corresponding to the noninteracting part of the Hamiltonian in
Eq.~(\ref{eq:hamiltonian}) can be written
$$\phi_{\bf n}({\bf x-R_m})=\prod\limits_{i=x,y,z}\phi_{n_i}^{(i)}(x_i-m_ia).$$ Here
the bold face vectors $\bf n$ and $\bf m$ are integer triplets
$(n_x,n_y,n_z)$ and $(m_x,m_y,m_z)$ which represent band indices
and lattice sites respectively, i.e.
$${\bf R}_{\bf m}=m_xa\hat{x}+m_ya\hat{y}+m_za\hat{z}.$$
These functions are to a good approximation described by localized harmonic
oscillator wave functions sketched in Fig.~\ref{fig:wavefuns}.
The completeness of the Wannier functions allows the field
operators to be expanded as
$$\hat{\psi}({\bf x})=\sum\limits_{\bf m}\sum\limits_{\bf n}\hat{d}_{\bf n}({\bf m}) \phi_{\bf n}({\bf x-R_m}).$$
The operators $\hat{d}^\dagger_{\bf n}({\bf m})$ and $\hat{d}_{\bf
n}({\bf m})$, which are the creation and annihilation operators of
bosons at site ${\bf m}$ and with band index ${\bf n}$, obey Bose
commutation relations $[\hat{d}_{\bf n}({\bf m}),\hat{d}_{\bf
n^\prime}^\dagger({\bf m}^\prime)]=\delta_{\bf n,\bf
n^\prime}\delta_{{\bf m},{\bf m}^\prime}.$ Ignoring all hopping
other than nearest neighbor hopping and all interactions other than
on-site interactions, the Hamiltonian in Eq.~(\ref{eq:hamiltonian})
can be written
\begin{widetext}
\begin{eqnarray}\label{eq:LatticeHamiltonian}
\hat{H}&\approx&\sum\limits_{\bf m}\sum\limits_{\bf n} E_{\bf n}({\bf
m})\hat{d}^\dagger_{\bf n}({\bf m})\hat{d}_{\bf n}({\bf
m})-\sum\limits_{i=x,y,z}\sum\limits_{\bf n} t_{\bf n
}^{(i)}\sum\limits_{\left<{\bf m,
m^\prime}\right>_i}\left[\hat{d}^\dagger_{\bf n}({\bf
m})\hat{d}_{\bf n}({\bf m^\prime})+\hat{d}^\dagger_{\bf n}({\bf
m}^\prime)\hat{d}_{\bf n}({\bf m})\right]\nonumber \\
&+&\frac{1}{2}\sum\limits_{{\bf
n}_{1},{\bf
n}_{2},{\bf
n}_{3},{\bf
n}_{4}}\sum\limits_{\bf m}U({\bf n}_{1},{\bf n}_{2},{\bf
n}_{3},{\bf n}_{4})\left[\hat{d}^\dagger_{{\bf n}_{1}}({\bf m})
\hat{d}^\dagger_{{\bf n}_{2}}({\bf m})\hat{d}_{{\bf n}_{3}}({\bf
m})\hat{d}_{{\bf n}_{4}}({\bf m})\right].
\end{eqnarray}
Here, the on-site interaction
energies are defined as
\begin{equation}
U({\bf n}_{1},{\bf n}_{2},{\bf n}_{3},{\bf n}_{4})\equiv\frac{4\pi
a_s\hbar^2}{m} \int d^3{\bf x}\, \phi^*_{{\bf n}_{1}}({\bf
x})\phi^*_{{\bf n}_{2}}({\bf x})\phi_{{\bf n}_{3}}({\bf
x})\phi_{{\bf n}_{4}}({\bf x}),
\label{eq:Unnnn}
\end{equation}
\end{widetext}
while the energies $E_{\bf n}({\bf m})$ and the hopping energies $t_{\bf n}^{(i)}$ are given by
\begin{equation}
E_{\bf n}({\bf m})\equiv\int d^3{\bf x}\, \phi^*_{\bf n}({\bf
x})\left(-\frac{\hbar^2}{2m}\nabla^2+V_O({\bf x})\right)\phi_{\bf
n}({\bf x})
\label{eq:En}
\end{equation}
\begin{equation}
t_{\bf n}^{(i)}\equiv\int dx_i\,
\phi^{(i)*}_{n_i}(x_i)\left(-\frac{\hbar^2}{2m}\frac{\partial^2}{\partial
x_i^2}+V_{0i}(x_i)\right) \phi_{n_i}^{(i)}({x_i}+a).
\label{eq:tn}
\end{equation}
Note that the energies $t^{(i)}_{\bf n}$ for hopping in the $x_i$-direction depend only on
the lattice depth $V_{0i}$ in the corresponding direction and the
$i$:th component $n_i$ of the band index ${\bf n}$.
The notation $\left<{\bf m},{\bf m}^\prime\right>_i$ in Eq.~(\ref{eq:LatticeHamiltonian}) indicates that the
sum should be carried out over nearest neighbor sites $\bf m$ and
$\bf m^\prime$ in the $x_i$-direction. One could for instance write,
$$\sum\limits_{\left<{\bf m},{\bf m}^\prime\right>_{y}}\equiv \sum\limits_{\bf m}\sum\limits_{{\bf m}^\prime}\delta_{m_x,m_x^\prime}\delta_{m_z,m_z^\prime}\delta_{m_y,m_y^\prime+1}$$

It is straight forward to numerically solve the noninteracting Schr{\"o}dinger equation and find the energies in expressions (\ref{eq:Unnnn})-(\ref{eq:tn}) above. In doing so, it is convenient to first switch to dimensionless units. Thus, we measure length in units of the inverse wave vector and potential depth in units of the
recoil energy $E_R$, i.e,
$\xi_i\equiv \frac{2\pi}{\lambda}x_i$ and $v_{0i}\equiv V_{0i}/E_R$,
with $E_R\equiv\frac{\hbar^2}{2m}\left(\frac{2\pi}{\lambda}\right)^2.$

\begin{figure}[ht]
{\includegraphics[width=8cm,clip]{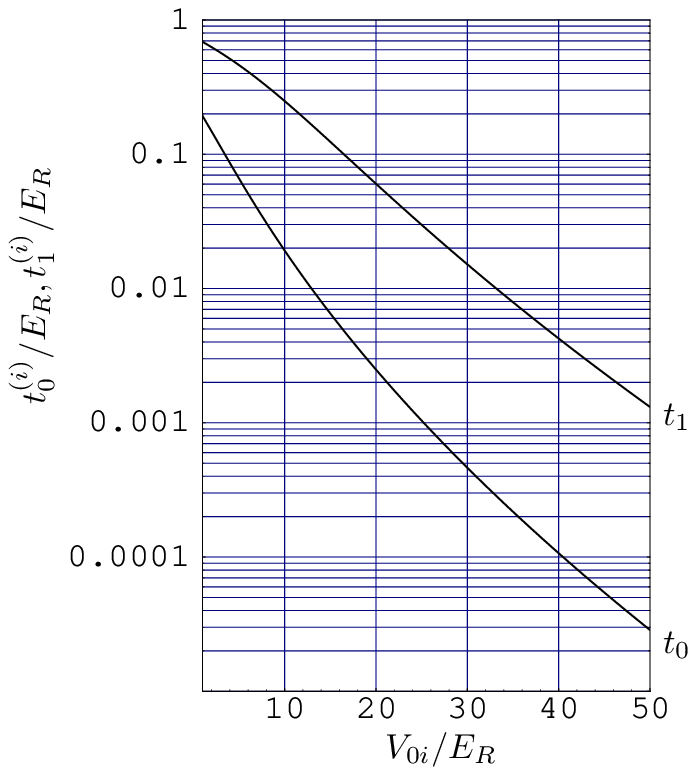}}
\caption{
 Hopping energies $t_1^{(i)}$ and
$t_0^{(i)}$ in units of the recoil energy as functions of lattice
depth $V_{0i}$ in the hopping direction. The upper line is the
hopping energy $t_1^{(i)}$ for atoms in the first Bloch band
hopping between nearest neighbor wells while the lower line,
$t_0^{(i)}$, corresponds to atoms in the zeroth band.
\label{fig:J12}}
\end{figure}

The hopping energies for the two lowest bands,
obtained from band-structure calculations, are shown in
Fig.~\ref{fig:J12} as functions of lattice depth.
To get the on-site interaction (Eq.~\ref{eq:Unnnn}) in a suitable form
to use later on in  the paper we define
dimensionless overlap integrals
\begin{equation}
O_{nn^\prime}(v)\equiv \sqrt{2\pi}\int d\xi\,
|\tilde{\phi}_n(v;\xi)|^2|\tilde{\phi}_{n^\prime}(v;\xi)|^2.
\label{eq:overlaps}
\end{equation} where dimensionless Wannier wave functions
$$\tilde{\phi}_n(v,\xi)\equiv
\sqrt{\frac{\lambda}{2\pi}}\phi_n(vE_R;\xi\lambda/2\pi)$$ have
been introduced. The dependence on $v$ in these functions is parametric, i.e.
$\phi_n(vE_R;\xi\lambda/2\pi)$ is the Wannier function
corresponding to the one-dimensional non-interacting problem with potential depth
$vE_R$. The variable $n$ denotes the band index.

\begin{figure}[t]
{\includegraphics[width=7.2cm]{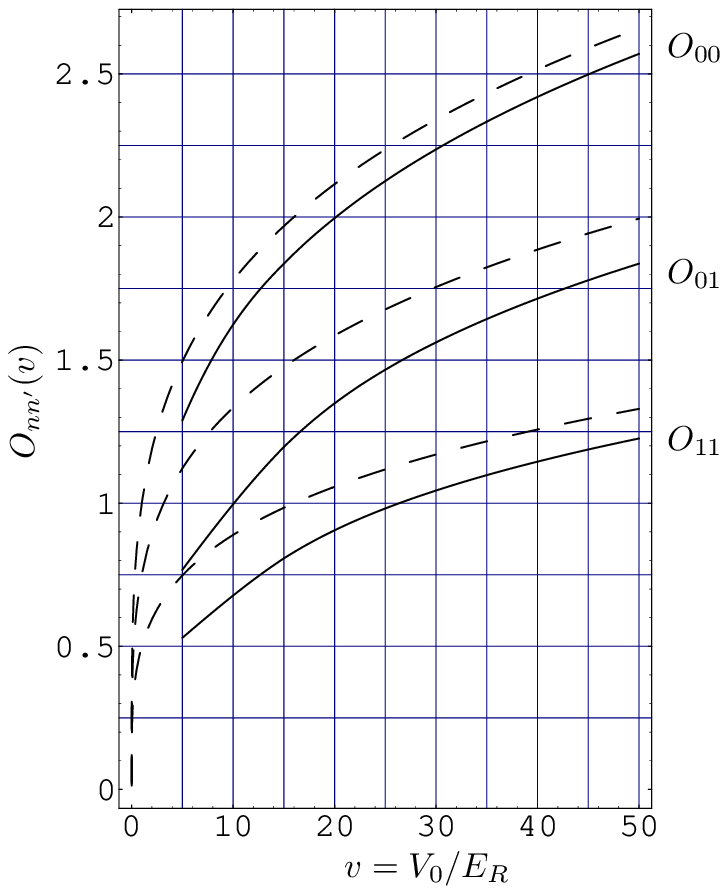}} \caption{ Overlap
integrals $O_{nn^\prime}(v)$ defined in Eq.~(\ref{eq:overlaps}).
Solid lines, from top to bottom, $O_{00},O_{01},O_{11}$, obtained
from numerical calculations. The dashed lines correspond to the
values in Eq.~(\ref{eq:hoapprox}) obtained by using harmonic
oscillator wave functions determined from the curvature of the
potential at the well bottom.
\label{fig:overlaps}}
\end{figure}

Approximating the Wannier functions with harmonic oscillator wave
functions corresponding to the curvature at the bottom of each well,
one finds approximate values for the overlap integrals
\begin{equation}
O^{\rm h.o.}_{nn^\prime}(v)=\frac{3^{nn^\prime}}{2^{n+n^\prime}}
v^{1/4}.
\label{eq:hoapprox}
\end{equation}
A comparison between these values for the overlap integrals and
those obtained from band-structure calculations is shown in
Fig.~\ref{fig:overlaps}.

\section{Effective Hamiltonians for atoms in the first excited band.}
\label{sec:effective_hams}
In this section we will focus on the meta-stable situation having
all atoms in the first Bloch band(s) of the lattice. It is quite
easy to achieve such a situation, i.e. consider an initial moment of
time when the optical lattice has been loaded with atoms in the
lowest Bloch band, ${\bf n} = (0,0,0)$. The anharmonicity of the
lattice well potential allows one to treat the vibrational degree of
freedom as a two level system. If one singles out, say, the
$x$-direction, then, by applying an appropriate vibrational
$\pi$-pulse, i.e.~``shaking" the lattice in this direction with a
frequency on resonance with the transition
$\hbar\omega=E_{(1,0,0)}-E_{(0,0,0)}$, the state can be inverted and
the atoms excited to states with band index ${\bf n}=(1,0,0)$. A
strong $\pi$-pulse can achieve this inversion in a time short
compared to the inter-well hopping time so that the dispersion of
the upper band is not an issue in the inversion process.  The
simplest starting state would be the Mott insulator state with one
boson per site in the lowest band. In a typical experimental setup,
the parabolic confining potential will cause the population of each
well to vary and the system will be in a state with regions of Mott
insulators with different filling factors. It is however easy to
confirm (by direct simulation) that even in this case, taking
interactions into account, a pulse shape can be tailored that will
invert the population simultaneously for regions with different
filling factors provided a deep enough lattice is used.

Another way of preparing the initial state is to use the method
recently demonstrated by Browaeys \emph {et al.}
\cite{Phillips:2005}. By loading a condensate into a moving 1D
lattice and applying a subsequent acceleration the condensate can
be prepared in the lowest energy state (quasimomentum $k=\pi/a$)
in the first Bloch band. The situation desired in this paper can
then be obtained by ramping up the lattice in the two remaining
(perpendicular) directions adiabatically. The natural question
regarding the lifetime of the resulting meta-stable state will be
considered in Sec.~\ref{sec:lifetime}.

The subsequent dynamics of atoms in the first band(s) is then
predominantly governed by some subset of the terms in
Eq.~(\ref{eq:LatticeHamiltonian}). This relevant subset will be
referred to as the effective Hamiltonian. We will treat three
different regimes of values for the lattice potentials $V_{0i}$
which lead to effective Hamiltonians with one, two, and three
flavors respectively. The three scenarios are:
\begin{enumerate}
\item $V_{0x}\ll V_{0y}, V_{0z}$ (1D, single flavor). \item
$V_{0x} = V_{0y} \ll V_{0z}$ (2D, two flavors). \item $V_{0x}=
V_{0y} = V_{0z}$ (3D, three flavors).
\end{enumerate}
As indicated, the number of particle
flavors as well as the dimensionalities in the effective Hamiltonians vary.

The reason for the different numbers of flavors becomes clear if one
considers the restrictions on the final states into which two atoms
may scatter due to the interatomic interaction; The presence or
absence of such states can be inferred from the presence or absence
of degenerate, or nearly degenerate, levels in the energy spectrum
of the noninteracting system. Take, for example the second scenario
above with $V_{0x}= V_{0y} \ll V_{0z}$ and all atoms initially in a
state with index ${\bf n}=(1,0,0)$, then, due to the on-site
inter-atomic interaction these atoms can scatter elastically into a
state with index ${\bf n}=(0,1,0)$ through a first order process the
connecting different degenerate states. Further, it is easy to show
that scattering resulting in states with other indices, for instance
${\bf n}=(0,0,1)$, is only possible through higher order processes
if energy (and also parity) is to be conserved and can safely be
ignored if the gas is dilute. Hence, the atoms can, at a formal
level, be divided into two flavors: an $X$ flavor corresponding to
atoms in ${\bf n}=(1,0,0)$ and a $Y$ flavor in ${\bf n}=(0,1,0)$. By
the same argument one can see how the one- and three-flavor
situations arise.

Apart from having different number of flavors the dimensionalities of the
effective Hamiltonians differ. To understand this consider again the
second case above, $V_{0x} = V_{0y} \ll V_{0z}$,
with particles in the excited bands ${\bf n}=(1,0,0)$ and ${\bf n}=(0,1,0)$
corresponding to $X$- and $Y$-flavors. For the $X$ flavor, hopping in the
$x$-direction has a matrix element $t_{1}^{(x)}(V_{0x})$ while hopping in the $y$- and $z$-
direction have matrix elements  $t_{0}^{(y)}(V_{0x})$ and  $t_{0}^{(z)}(V_{0z})$ respectively.
Looking at Fig.~\ref{fig:J12} it is then clear that, to a good approximation, the $X$
particles can only hop in the $x$-direction while hopping in the $y$- and $z$-directions
is strongly (exponentially) suppressed. Similarly the $Y$ particles can
only hop in the $y$-direction and all hopping occurs only in the $x-y$ plane, hence the 2D character.
A similar argument holds for the three flavor case where in addition to the $X$ and $Y$ particles, there are
$Z$- particles hopping in the $z$-direction.

The effective Hamiltonians also contain terms arising from the
on-site interaction. Apart from the terms that repel atoms from each
other, the symmetry of the on-site interaction allows, say, two $X$
particles moving in the $x$-direction to collide and convert into
two $Y$ particles which thereafter move off in the $y$-direction.
The time reversed process can of course also occur. Thus, the number
of particles of each flavor is not conserved and there is a pairwise
exchange of particles of different flavors.

\begin{figure}[ht]
\begin{center}
\hspace{-2cm}\includegraphics[angle=-90,width=6.5cm]{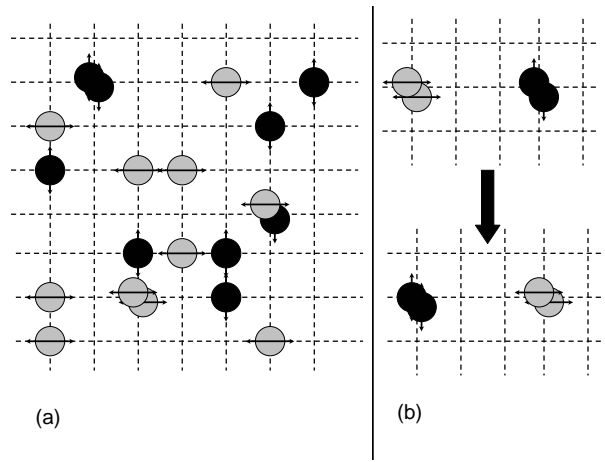}
\vspace*{-3cm}
\caption{(a) The Hamiltonian in Eq.~(\ref{eq:2dham}) describes a 2D system where
atoms, which can formally be thought of as having two different flavors (same type of atoms but
in different localized on-site orbitals), hop around subject to on-site
repulsive interactions. One flavor, the $X$-flavor, can only hop in the $x$-direction whereas the
other, $Y$-flavor, can only hop in the $y$-direction.
(b) Conversion process. The conversion term in Eq.~(\ref{eq:2dham}) takes two $X$-atoms on the same
lattice site and turns them into two $Y$-atoms or vice versa.\label{fig:2Dillustr}}
\end{center}
\end{figure}

The anisotropy of hopping and the flavor conversion process is schematically
depicted for the 2D
(two flavors)
case in Fig.~\ref{fig:2Dillustr}. Particles of $X$ flavor are shown
in gray while the $Y$ flavor is drawn in black.

Below, we give the effective
Hamiltonians for all three different cases listed above.

\subsection{1D Hamiltonian, single flavor, {$(V_{0x} \ll V_{0y}=V_{0z})$}}
The first case to be considered
is when $V_{0x} \ll V_{0y}=V_{0z}$ and only states with band index
${\bf n}=(1,0,0)$ (and possibly some residual atoms in ${\bf
n}=(0,0,0)$) are occupied. In anticipation of the other effective
Hamiltonians it is convenient to introduce for the $X$ flavor
the creation and destruction operators $\hat{X}^\dagger$ and $\hat{X}$, \emph{i.e.}:
\begin{eqnarray}
\hat X_{\bf m}&\equiv& \hat d_{(1,0,0)}({\bf m}),\,\,\hat X_{\bf m}^\dagger \equiv \hat d_{(1,0,0)}^\dagger({\bf m})\nonumber \\
\hat{n}_{\bf m}^{(x)}&\equiv&\hat X_{\bf m}^\dagger\hat X_{\bf
m},\,\,\hat{n}_{\bf m}^{(0)}\equiv \hat d_{(0,0,0)}^\dagger({\bf
m})\hat d_{(0,0,0)}({\bf m}).
\end{eqnarray}
The effective Hamiltonian is then essentially that of a quasi
one-dimensional bosonic Hubbard model
\begin{eqnarray}
\label{eq:1DLatticeHamiltonian} H_{1D}&=&\sum\limits_{\bf
m}\hat{n}_{\bf m}^{(x)}\left(E_{x}({\bf m})+U_{0x}\hat{n}_{\bf
m}^{(0)}+\frac{U_{xx}}{2}[\hat{n}_{\bf m}^{(x)}-1]\right)\nonumber \\
&-&t\sum\limits_{\left<{\bf m,
m^\prime}\right>_x}\left[\hat{X}^\dagger_{\bf m}\hat{X}_{{\bf m}^\prime}+{\rm h.c.}\right]\nonumber
\end{eqnarray}
The energies $t\equiv t_1^{(x)}$, $U_{0x}$, and $U_{xx}$ arise from
the inter-well tunneling and the inter-atomic interaction
respectively. The presence of atoms residing in the lowest band
leads to an additional effective on-site energy and can be
absorbed in the on-site energies $E_{x}({\bf m})$, $(s=x,y,z)$.
Although this single-flavor model is equivalent to a single-flavor
model in the zeroth Bloch band, the additional random on-site
potential resulting from residual atoms could be exploited in the
study of the disordered Bose-Hubbard system.

The parameters entering the Hamiltonian
(\ref{eq:1DLatticeHamiltonian}) are conveniently expressed as
\begin{equation}
U_{xx}=2\sqrt{2\pi}E_R\left(\frac{a_s}{a}\right)O_{00}(v_{0y})O_{00}(v_{0z})O_{11}(v_{0x}),
\label{eq:Uaa}
\end{equation}
\begin{equation}
U_{0x}=2U_{xx}O_{01}(v_{0x})/O_{11}(v_{0x}).
\end{equation}
\subsection{2D Hamiltonian, two flavors, $(V_{0x}=V_{0y}\ll V_{0z})$}
To simplify the notation for the case $V_{0x}=V_{0y} \ll V_{0z}$, we
introduce new letters for the creation/annihilation operators $\hat
Y_{\bf m}\equiv \hat d_{(0,1,0)}({\bf m})$, $\hat{n}_{\bf
m}^{(y)}\equiv\hat Y_{\bf m}^\dagger\hat Y_{\bf m}$. Then, the
Hamiltonian governing atoms in the excited bands becomes
\begin{eqnarray}
H_{2D}=\sum\limits_{s=x,y}\sum\limits_{\bf m}E_{s}({\bf
m})\hat{n}^{(s)}_{\bf m}
+\sum\limits_{s=x,y}\frac{U_{ss}}{2}\sum\limits_{ \bf
m}\hat{n}^{(s)}_{\bf
m}[\hat{n}^{(s)}_{\bf m}-1]\nonumber\\
-t\sum\limits_{\left<{\bf m,
m^\prime}\right>_y}\left[\hat{Y}^\dagger_{\bf m}\hat{Y}_{\bf
m^\prime}+{\rm h.c.}\right]-t\sum\limits_{\left<{\bf m,
m^\prime}\right>_x}\left[\hat{X}^\dagger_{\bf m}\hat{X}_{\bf
m^\prime}+{\rm h.c.}\right]\nonumber \\
+U_{xy}\sum\limits_{{\bf m}}\hat{n}^{(x)}_{\bf
m}\hat{n}^{(y)}_{\bf m}+\frac{U_{xy}}{2}\sum_{\bf
m}[\hat{X}^\dagger_{\bf m}\hat{X}^\dagger_{\bf m}\hat{Y}_{\bf
m}\hat{Y}_{\bf m}+{\rm h.c.}].\nonumber\\ \label{eq:2dham}
\end{eqnarray}
Again, the energy $U_{xy}$ arises from the interatomic interaction
and depends on the lattice depth. Note that this two flavor
bosonic Hubbard Hamiltonian differs in an important aspect from
previously studied two flavor systems: the presence of the last
term that mixes the two flavors. Hence, the inter-atomic
interaction leads to a ``Josephson term" that allows for the
conversion of two $X$-atoms into two $Y$-atoms and vice versa. The
coefficients $U_{yy}=U_{xx}$ are given by the same expression as
in the 1D case while
\begin{equation}
U_{xy}=2\sqrt{2\pi}E_R\left(\frac{a_s}{a}\right)O_{00}(v_{0z})O_{01}(v_{0x})^2.\label{eq:Uabexp}
\end{equation}
Figure~\ref{fig:2Dillustr} illustrates the dynamics in the 2D
(two flavors)
situation.

\subsection{3D Hamiltonian, three flavors, $(V_{0x}=V_{0y}=V_{0z})$}
The generalization of the above
Hamiltonian to the case when $V_{0x}=V_{0y}=V_{0z}$ is straight
forward. Introducing a third flavor $\hat Z_{\bf m}\equiv\hat
d_{(0,0,1)}({\bf m})$, $\hat{n}_{\bf m}^{(z)}\equiv\hat Z_{\bf
m}^\dagger\hat Z_{\bf m}$, one may write an effective Hamiltonian
as
\begin{eqnarray}
H_{3D}&=&\sum\limits_{s=x,y,z}\sum\limits_{\bf m}\left(E_{s}({\bf
m})\hat{n}^{(s)}_{\bf m} +\frac{U_{ss}}{2}\hat{n}^{(s)}_{\bf
m}[\hat{n}^{(s)}_{\bf m}-1]\right)\nonumber \\
&+&\sum\limits_{s\neq s^\prime}\sum\limits_{{\bf
m}}U_{ss^\prime}\left(\hat{n}^{(s)}_{\bf
m}\hat{n}^{(s^\prime)}_{\bf m}+\frac{1}{2}[\hat{s}^\dagger_{\bf
m}\hat{s}^\dagger_{\bf m}\hat{s^\prime}_{\bf m}\hat{s^\prime}_{\bf
m}+{\rm h.c.}]\right)\nonumber\\
&-&t\sum\limits_{s=X,Y,Z}\sum\limits_{\left<{\bf m,
m^\prime}\right>_s}\left[\hat{s}^\dagger_{\bf m}\hat{s}_{\bf
m^\prime}+{\rm h.c.}\right].\label{eq:3dham}
\end{eqnarray}
Here
$U_{ss^\prime}=\delta_{ss^\prime}U_{xx}+(1-\delta_{ss^\prime})U_{xy}$
with $U_{xx}$ and $U_{xy}$ given by
(\ref{eq:Uaa}) and (\ref{eq:Uabexp}) with
$v_{0x}=v_{0y}=v_{0z}$.

\subsection{$Z_2$ Gauge symmetry}
\label{sec:Z2}
Because of overall number conservation the Hamiltonian has the usual global $U(1)$ symmetry.
However, because the flavor conversion occurs pairwise and locally (\emph{i.e.}, on site),
the Hamiltonians described above also exhibit an infinite number of $Z_2$ gauge symmetries corresponding to conservation modulo 2 of the number
of $X$ particles in any given column of the lattice running in the $x$-direction (and similarly for $Y$ and $Z$ particles).
These symmetries correspond to invariance under each of the transformations
$$U^{(m_y,m_z)}_X=\exp\left[i\pi\sum\limits_{m_x}\hat{X}_{(m_x,m_y,m_z)}^\dagger\hat{X}_{(m_x,m_y,m_z)}\right]$$
$$U^{(m_x,m_z)}_Y=\exp\left[i\pi\sum\limits_{m_y}\hat{Y}_{(m_x,m_y,m_z)}^\dagger\hat{Y}_{(m_x,m_y,m_z)}\right]$$
$$U^{(m_x,m_y)}_Z=\exp\left[i\pi\sum\limits_{m_z}\hat{Z}_{(m_x,m_y,m_z)}^\dagger\hat{Z}_{(m_x,m_y,m_z)}\right],$$
where the integer pair $(m_i,m_j)$ in the superscript of each $U$
determine the location of a column. The first transformation for
example takes $\hat{X}_{(m_x,m_y,m_z)}\rightarrow
-\hat{X}_{(m_x,m_y,m_z)}$ for all $m_x$ in the column specified by
$m_y$ and $m_z$. Since $\hat{X}^\dagger$ and $\hat{X}$ operators
always appear pairwise, the Hamiltonian is invariant under this
class of $Z_2$ transformations. These $Z_2$ symmetries are in a
sense intermediate between local and global. While the number of
such symmetries is infinite (in the thermodynamic limit) it is of
course sub extensive and thus not large enough to fully constrain
the system (or to make it integrable for example).
As mentioned in the introduction, such symmetries have been found in
certain frustrated spin
models~\cite{Moore1,Moore2,Nussinov1,Nussinov2} and in a `bose
metal' model~\cite{Paramekanti} and are known to cause dimensional
reduction in some cases.~\cite{Moore1,Moore2,Nussinov2} Because
introducing a defect across which the sign of the $Z_2$ order
parameter changes along any given single column costs only finite
energy, the system will, like the 1D Ising model, disorder at any
finite temperature thereby restoring the $Z_2$ symmetry. We will see
below how this reduced dimensionality physics appears in a simple
way in this system.

\section{Mean field theory phase diagrams for the effective Hamiltonians}
\label{sec:MFT}
Having derived effective Hamiltonians in one, two, and three
dimensions, we turn now to the investigation of their ground states.
The 1D, single flavor Hamiltonian has been extensively studied (see
for instance Ref.~\cite{Zwerger:2002} and references therein) and
needs no further discussion here. The other two Hamiltonians in
Eqs.~(\ref{eq:2dham}) and (\ref{eq:3dham}) deserve some attention
though.

\subsection{Phase diagram 2D, two flavors}
The 2D Hamiltonian (\ref{eq:2dham}) is a two flavor bosonic Hubbard
Hamiltonian, a system that has recently received much attention and
shown to have a rich phase
diagram~\cite{Chen:2003,Kuklov:2002,Kuklov:2003,Altman:2003,Duan:2002,Ziegler:2003}.
In this section we will investigate the ground state of the
Hamiltonian in Eq.~(\ref{eq:2dham}) using simple mean field theory.
The Hamiltonian~(\ref{eq:2dham}) differs from those previously
studied in two aspects: the presence of pairwise inter flavor mixing
and the anisotropic tunneling.

We follow here the method suggested in Ref.~\cite{Sheshadri:1993}
(see also Refs.~\cite{Sachdev, Oosten1:2001}). We consider the
possibility that the global $U(1)$ and columnar $Z_2$-symmetries
discussed in section~\ref{sec:Z2} are spontaneously broken by
introducing complex scalar columnar order parameter fields
$\psi_x(m_y)$ and $\psi_y(m_x)$, \emph{i.e.} one for each $x$-column and
one for each $y$-column. These fields should then satisfy the self
consistency conditions
\begin{equation}
\psi_x(m_y)=\left<\hat{X}_{(m_x,m_y)}\right>
\label{eq:scons1}
\end{equation}
for all $m_x$ in the $x$-column specified by $m_y$, and,
\begin{equation}
\psi_y(m_x)=\left<\hat{Y}_{(m_x,m_y)}\right>
\label{eq:scons2}
\end{equation}
for each $m_y$ in the $y$-column specified by $m_x$. For simplicity
we omit in this discussion of the 2D
(two flavors)
case the $z$ component $m_z$ of
the position vector ${\bf m}$.  The possibility that fluctuations
restore the symmetry will be discussed further below.

Mean field theory results from decoupling sites in the same column by neglecting
fluctuations in the kinetic energy. For instance, for the $x$-column specified by
a particular value of $m_y$ one has
\begin{widetext}
\begin{eqnarray}
\hat{X}^\dagger_{(m_x,m_y)}\hat{X}_{(
m_x+1,m_y)}&=&(\hat{X}^\dagger_{(m_x,m_y)}-\psi_x^*(m_y)+\psi_x^*(m_y))(\hat{X}_{(m_x+1,m_y)}-\psi_x(m_y)+\psi_x(m_y))\nonumber \\
&\approx& \psi_x(m_y)\hat{X}^\dagger_{(m_x,m_y)}+\psi_x^*(m_y)\hat{X}_{(m_x+1,m_y)}-|\psi_x(m_y)|^2, \nonumber
\end{eqnarray}
Thus the sites along each column decouple. Doing the same for the
$Y$:s and writing the Hamiltonian in dimensionless form where all
energies are scaled by $U_{xx}$, \emph{i.e.}, $h_{2d}\equiv H_{2D}/U_{xx}$,
$\tilde{t}\equiv t/U_{xx}$ and $\tilde{U}_{xy}\equiv U_{xy}/U_{xx}$,
we obtain $h_{2D}\approx\sum\limits_{\bf m}h_{2d}^{MF}({\bf
m};\psi_x(m_y),\psi_y(m_x)).$ Here, the on-site mean field
Hamiltonians are given by
\begin{eqnarray}
h_{2d}^{MF}({\bf m};\psi_x(m_y),\psi_y(m_x))&=&-2\tilde{t}\left[\psi_x(m_y)\hat{X}^\dagger_{\bf m}+\psi_x^*(m_y)\hat{X}_{\bf m}\right]
-2\tilde{t}\left[\psi_y(m_x)\hat{Y}^\dagger_{\bf m}+\psi_y^*(m_x)\hat{Y}_{\bf m}\right]+\tilde{U}_{xy}\hat{n}^{(x)}_{\bf m}\hat{n}^{(y)}_{\bf m}\nonumber \\
&+&\sum\limits_{s=x,y}\left(\frac{1}{2}\hat{n}^{(s)}_{\bf m}[\hat{n}^{(s)}_{\bf m}-1]-\tilde{\mu}\hat{n}^{(s)}_{\bf m}+2\tilde{t}|\psi_s({\bf m})|^2\right)+\frac{\tilde{U}_{xy}}{2}[\hat{X}^\dagger_{\bf m}\hat{X}^\dagger_{\bf m}\hat{Y}_{\bf m}\hat{Y}_{\bf m}+
\hat{Y}^\dagger_{\bf m}\hat{Y}^\dagger_{\bf m}\hat{X}_{\bf m}\hat{X}_{\bf m}],\nonumber\\
\label{eq:2dham2}
\end{eqnarray}
\end{widetext}
where $\tilde{\mu}\equiv\mu/U_{xx}$ serves as a common chemical potential. The on site Hamiltonians satisfy
the eigenvalue relations
$$h_{2d}^{MF}({\bf m};\psi_x,\psi_y)\left|\epsilon_{n}(\psi_x,\psi_y)\right>=\epsilon_{n}(\psi_x,\psi_y)\left|\epsilon_{n}(\psi_x,\psi_y)\right>$$
for two arbitrary complex fields. An eigenstate of the full mean field hamiltonian can be written as a product
state of such eigenstates
$$\left|\Psi\right>=\prod\limits_{\bf m}\left|\epsilon_{n_{\bf m}}(\psi_x(m_y),\psi_y(m_x))\right>$$
where the fields satisfy the self consistency conditions in Eqs.~(\ref{eq:scons1})-(\ref{eq:scons2}).
The mean field ground state is obtained by globally minimizing the
energy
$$E=\sum\limits_{\bf m}\epsilon_{n_{\bf m}}(\psi_x(m_y),\psi_y(m_x))$$ with respect to the fields and the set of eigenstates $\{n_{\bf m}\}$. This is most easily done by numerical diagonalization in a truncated Hilbert space where
each site can hold at most a total of $N_{\rm max}$ atoms. Since
\begin{eqnarray}
\min\limits_{\left[n_{\bf m},\psi_x(m_y),\psi_y(m_x)\right]}&&\left(\sum\limits_{\bf m}\epsilon_{n_{\bf m}}(\psi_x(m_y),\psi_y(m_x))\right)\nonumber \\
&\ge&\sum\limits_{\bf m}\min\limits_{\left[n,\psi_x,\psi_y\right]}\epsilon_{n}(\psi_x,\psi_y)\nonumber
\end{eqnarray}
\begin{figure}[b]
\includegraphics[width=8cm,clip]{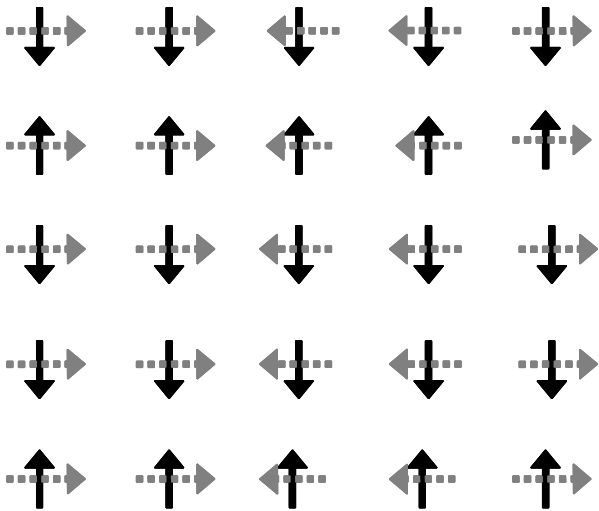}
\caption{
Columnar phase ordering in 2d superfluid phase. The directions of the arrows correspond to
the phase angles $\phi_x(m_y)$ and $\phi_y(m_x)$ of the order parameter fields
$\psi_x(m_y)=|\psi_x(m_y)|e^{i\phi_x(m_y)}$ and $\psi_y(m_x)=|\psi_y(m_x)|e^{i\phi_y(m_x)}$. Solid arrows
correspond to $\phi_x$ and dashed to $\phi_y$. \label{fig:2dphases}}
\end{figure}
it is enough to minimize the ground state energy of a single site
with respect to the fields and then find the largest manifold of
states compatible with having columnar order parameters fields.
Carrying out this scheme reveals two different scenarios for the
minimum of each on-site energy $\epsilon_{n}(\psi_x,\psi_y)$; either
$\psi_x=\psi_y=0$ and $n_x+n_y$ is integer (incompressible), or,
$|\psi_x|=|\psi_y|\neq 0$ (compressible). The former case
corresponds to a Mott insulating state while the latter suggests a
superfluid phase.

Due to the positivity of $U_{xy}$ the last term in
the mean field Hamiltonian is minimal whenever
$\psi_x$ and $\psi_y$, on the same site, have a phase difference of $\pm\pi/2$.

For the (mean field) ground state manifold we must have in this
phase $|\psi_x(m_y)|=|\psi_y(m_x)|$ for all $x-$ and $y-$ column
order parameters. Requiring the phases of all $\psi_x$ in each
$x-$ and all $\psi_y$ in the $y-$ columns to be the same while
fixing the relative phase between $\psi_x$ and $\psi_y$ to $\pm
\pi/2$ results in configurations as the one shown in
Fig.~\ref{fig:2dphases}. Here the phases of $\psi_x$ and $\psi_y$
are shown represented as arrows (planar spins). The direction of
the arrows defining the angle. Clearly, this phase shows a
breaking of the global $U(1)$ symmetry. The meaning of the
quasi-local nature of the $Z_2$ symmetries discussed above becomes
clear. Although the phases of $\psi_x$ in each $x-$ column are the
same there is no energy cost associated with flipping all the
spins $x$-spins in a single column or all the $y$-spins in a
$y$-column. The ordering between different columns is thus
nematic.

One should note here, that since the only energy cost associated
with flipping a single spin, say an $x$-spin in an $x-$column, is
given by the states of the neighboring $x$ spins in the column, the
situation is essentially that of a 1D Ising model along each column.
Hence, at any finite temperature, domains of flipped spins will
proliferate and the $Z_2$ symmetries will be restored. This
essentially one dimensional behavior is an example of the
dimensional reduction mentioned above.

The model under consideration is highly anisotropic. Mainly since
$X$ particles can only hop in the $x$-direction, it seems to be
impossible to develop phase coherence among $X$ particles in
different $x$-columns (and similarly for the other flavors).
Suppose however that,
as discussed above,
the flavor exchange
interaction term causes the relative phase of two flavors, say $X$
and $Y$, to lock together so that $\hat{Y}^\dagger\hat{X}$
condenses
$$ \psi\equiv\left<\hat{Y}^\dagger\hat{X}\right>\neq 0.$$
In this case the mean field decomposition of the exchange
interaction yields terms of the form
$$V\sim \psi\hat{Y}^\dagger\hat{X}+\psi^*\hat{X}^\dagger\hat{Y}$$
which permit individual particles to change flavor and hence phase
coherence can freely propagate in all directions throughout the
lattice via a kind of `Andreev' process (\emph{i.e.} self-energy
off-diagonal in flavor index) in which an $X$ particle can turn
into a $Y$ particle when it needs to travel in the $y$ direction.

To understand this isotropic superfluid phase, it is convenient to
consider a phase only representation with compact phase variables
on each site $\hat{X}_{\bf m}\rightarrow e^{-i\varphi_{\bf m}^x}$
and $\hat{Y}_{\bf m}\rightarrow e^{-i\varphi^y_{\bf m}}$. The
flavor exchange (`Josephson') term then becomes (for the 2D
(two flavors) case)
$$V=\tilde U_{xy}\sum\limits_{\bf m}\cos(2[\varphi_{\bf m}^x-\varphi_{\bf m}^y]).$$
Defining $\varphi_{\bf m}^\pm\equiv \varphi_{\bf
m}^x\pm\varphi_{\bf m}^y$ we have
$$V=\tilde U_{xy}\sum\limits_{\bf m}\cos(2\varphi_{\bf m}^-).$$
Assuming that the relative phase of the condensates is locked
together by this Josephson term is equivalent to assuming that
(for $U_{xy}>0$) the fluctuations of $\varphi^-$ away from the
ground state value $\pi/2$ (or its equivalent $-\pi/2$ under the
$Z_2$ gauge symmetry) are massive and can be ignored. Thus we
obtain $\varphi_{\bf m}^{(x,y)}={\varphi^+_{\bf
m}}/{2}\pm{\pi}/{4}.$ The continuum limit of the anisotropic
kinetic energy
$T=\left(\partial_x\varphi^x\right)^2+\left(\partial_y\varphi^y\right)^2$
then becomes $T\sim
\left((\partial_x\varphi^+)^2+(\partial_y\varphi^+)^2\right)$ and
we immediately see that the anisotropy has effectively disappeared
at long wavelengths and we have a superfluid.

\begin{figure}[t]
\includegraphics[width=8cm]{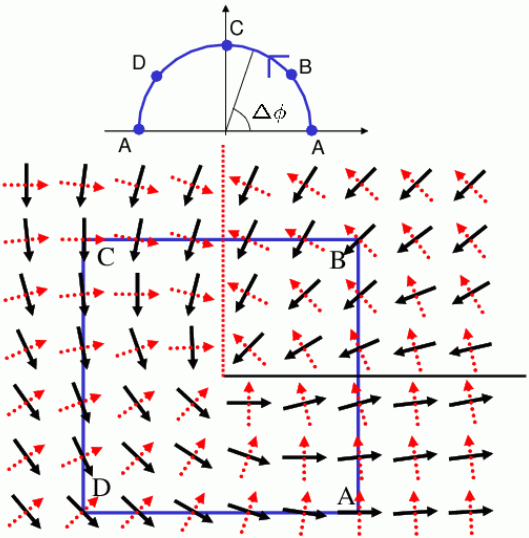}
\caption{(color online) Phase configuration for $\psi_x$ (black
solid arrows) and $\psi_y$ (red dashed arrows) containing a half
vortex.  Notice the branch cuts indicated by the long solid black
and dashed red lines.  The $Z_2$ symmetry means that these branch
cuts have zero `string tension' and contribute only a finite core
energy to the vortex.
The half winding number can be seen by going around a loop A-B-C-D-A
and calculating the total phase twist $\Delta \phi$. This twist is calculated by summing
the changes in $\varphi^y$ when going vertically and $\varphi^x$ when going horizontally.
\label{fig:half-vortex}}
\end{figure}

A vortex
configuration in $\varphi^+$ can be viewed as a bound state of two
half vortices in the $\varphi^x$ and $\varphi^y$ fields.  The
columnar $Z_2$ symmetry allows the $\psi_x$ field to have a phase
jump of $\pi$ across a cut parallel to the $x$ axis and similarly
for $\psi_y$. Thus half vortices are permitted.  If the two order
parameter phases are locked together ($\varphi^-$ fluctuations are
massive) then the two half vortices are confined to each other as
shown in Fig.~(\ref{fig:half-vortex}).  Such a vortex has an
energy which scales (as usual) only logarithmically with system
size, despite the semi-infinite branch cut ($\pi$ phase jump) of
$\psi_x$ running horizontally out to the right from the vortex
center and of the similar branch cut in $\psi_y$ running
vertically out above the vortex center.
To see that such a vortex is topologically well defined despite the $Z_2$ symmetry
one can consider a loop around the vortex core as shown in Fig.~(\ref{fig:half-vortex}).
In going around the loop we add up the phase twist $\Delta \phi$ and map onto
the complex plane. To calculate $\Delta \phi$ along the loop the changes in $\varphi^y$ has
to be added when going vertically and the changes in $\varphi^x$ when going horizontally.
The net results is that going once around the vortex core the phase winds by $\pi$.
If one applies a $\pi$-flip in all the $\varphi^x$ ($\varphi^y$) phases in any row (column)
the mapping onto the complex plane will remain invariant.

The Mott insulating states, having integer
number of atoms in each well,
are best characterized by the
$\tilde{t}=0$ eigenstates. These are product states
$$\left|\Psi_0(\tilde{t}=0)\right>=\prod\limits_{\bf
m}\left|\psi_{Ni}({\bf m})\right>,$$ where
$h_{2D}^{MF}(0,0)\left|\psi_{Ni}\right>=\epsilon_{Ni}\left|\psi_{Ni}\right>$
and the integer $N$ is the total number of particles $N=n_x+n_y$ in
each well. The index $i$ runs from $0$ to $N$ for each $N$ and for
the three lowest values of $N$ the eigenstates are

$\left|\psi_{00}\right>=\left|0\right>, \epsilon_{00}=0.$

$\left|\psi_{10}\right>=\left|1_x,0_y\right>,
\epsilon_{10}=-\tilde{\mu}.$

$\left|\psi_{11}\right>=\left|0_x,1_y\right>,
\epsilon_{11}=-\tilde{\mu}.$

$\left|\psi_{20}\right>=\left|1_x,1_y\right>,
\epsilon_{20}=-2\tilde{\mu}+\tilde{U}_{xy}.$

$\left|\psi_{21}\right>=\frac{1}{\sqrt{2}}\left(\left|2_x,0_y\right>+\left|0_x,2_y\right>\right),
\epsilon_{21}=1-\tilde{U}_{xy}-2\tilde{\mu}.$

$\left|\psi_{22}\right>=\frac{1}{\sqrt{2}}\left(\left|2_x,0_y\right>-\left|0_x,2_y\right>\right),
\epsilon_{22}=1+\tilde{U}_{xy}-2\tilde{\mu}.$

\begin{figure}[t]
\includegraphics[width=8cm]{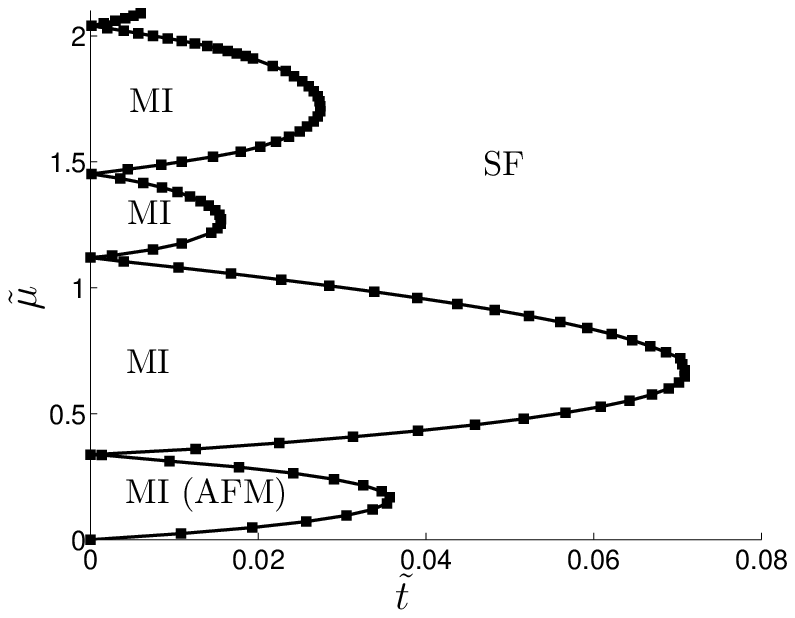}
\caption{
Mean field ground state phase diagram for the 2D
(two flavors) Hamiltonian in
Eq.~\ref{eq:2dham2} in the plane of $\tilde{\mu}$, the scaled
chemical potential $\mu/U_{xx}$, and $\tilde{t}=t/U_{xx}$, the
scaled hopping energy. Here calculated for the experimentally
relevant ratio $\tilde{U}_{xy}\equiv{U_{xy}/U_{xx}=1/3}$ using a
truncated Hilbert space with at most 10 particles per site. Lobes of
Mott-insulating states, of successively increasing integer filling
factor with increasing chemical potential $\tilde{\mu}$, are
surrounded by a superfluid phase. The superfluid phase is
characterized by columnar order parameter fields $\psi_x(m_y)$ and
$\psi_y(m_x)$, one for each $x$- and $y$- column respectively.  All
$\psi_x$ and $\psi_y$ have equal, nonzero, magnitudes while their
relative phases are either $0$ or $\pi$. (see
Fig.~\ref{fig:2dphases}).
\label{fig:phasediag2D}}
\end{figure}

In Fig.~\ref{fig:phasediag2D} the mean field phase diagram has been
drawn for the physically relevant value $\tilde{U}_{xy}=1/3$ which
is characteristic for the proposed setup. The lobes marked MI
correspond to incompressible Mott insulating phases with integer
filling factors. The remaining part of the diagram, marked SF,
corresponds to a superfluid phase with the columnar nematic ordering
(see Fig.~\ref{fig:2dphases}) discussed above. Considering the $t=0$
eigenstates above two things become clear. Trivially, if
$\tilde{U}_{xy}\rightarrow 0$ the lowest lobe, and all other odd
filling lobes, vanish and the model reduces to two noninteracting
single flavor models as expected. Secondly, at $\tilde{U}_{xy}=0.5$
there is a level crossing between $\left|\psi_{20}\right>$ and
$\left|\psi_{21}\right>$. It follows that the size of the lowest odd
filling lobes increases with increasing values of $\tilde{U}_{xy}$
up until $\tilde{U}_{xy}=0.5$ after which it starts to decrease
again.

By considering fluctuation effects higher order in the tunneling
amplitude, we can  demonstrate that the permutational symmetry
between the $X$- and $Y$- flavors can be broken in the Mott
insulator phase. In the absence of tunneling, the single-particle
states $\left|\psi_{10}\right>$ and $\left|\psi_{11}\right>$ are
degenerate. Taking tunneling into account breaks this degeneracy and
to second order in $\tilde{t}$ (using for instance the
Schrieffer-Wolff transformation~\cite{Schrieffer:1966}) an effective
(pseudo) spin-$\frac{1}{2}$ Hamiltonian for the interaction between
neighboring sites can be found
\begin{equation}
H_{eff}=-J_{eff}\sum\limits_{\left<{\bf m},{\bf
m^\prime}\right>}\hat{\sigma}^{(z)}_{\bf m}\hat{\sigma}^{(z)}_{\bf
m^\prime}
\end{equation}
The up- and down- states of the pseudo spin operators
$\hat{\sigma}^{(z)}_{\bf m}$ correspond to the site ${\bf m}$ being
occupied by one $X$-atom or one $Y$-atom respectively. The effective magnetic interaction is
$$J_{eff}=\frac{\tilde{t}^2(\tilde{U}_{xy}^2+2\tilde{U}_{xy}-1)}{\tilde{U}_{xy}(1-\tilde{U}_{xy}^2)}.$$
There is a critical value of the inter flavor interaction
$\tilde{U}_{xy}^{c}=\sqrt{2}-1\approx 0.414$ for which $J_{eff}$
vanishes. For $\tilde{U}_{xy}>\tilde{U}_{xy}^{c}$, the system is
ferromagnetic and spontaneously favors one flavor over the other.
For $\tilde{U}_{xy}<\tilde{U}_{xy}^{c}$ the system is
anti-ferromagnetic and favors an ordering with $X$- and $Y$-atoms on
alternating sites. Thus, we conclude that at integer filling factor
the permutational symmetry  between $X$ and $Y$ flavors, (or
equivalently, the cubic symmetry of the underlying lattice) is
always broken in the mean field ground state. Further, in the
anti-ferromagnetic state, sublattice (\emph{i.e.} translation) symmetry is
broken as well.

\begin{figure}[t]
\includegraphics[width=7cm]{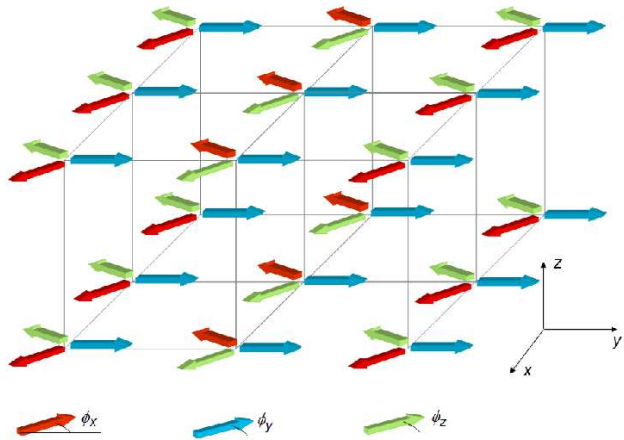}
\caption{
(Color online) Phase ordering in 3d superfluid phase. The directions
of the arrows correspond to the phase angles $\phi_x(m_y,m_z)$,
$\phi_y(m_x,m_z)$ and $\phi_z(m_x,m_y)$ of the order parameter
fields $\psi_x(m_y,m_z)=|\psi_x(m_y,m_z)|e^{i\phi_x(m_y,m_z)}$,
$\psi_y(m_x,m_z)=|\psi_y(m_x,m_z)|e^{i\phi_y(m_x,m_z)}$ and
$\psi_z(m_x,m_y)=|\psi_y(m_x,m_y)|e^{i\phi_y(m_x,m_y)}$. As in the
2D (two flavors)
case the underlying symmetry of the Hamiltonian allows for
flipping say all the $\phi_x$ along any $x$-column by $\pi$ to
obtain another ground state configuration. In addition to the ground
state degeneracy obtained from such operations, an accidental
degeneracy associated with parallel planes of different chirality is
present. In this figure the middle $x-z$ plane has a different
chirality than the other two $x-z$ planes.\label{fig:3dphases}}
\end{figure}
\begin{figure}[t]
\includegraphics[width=8cm]{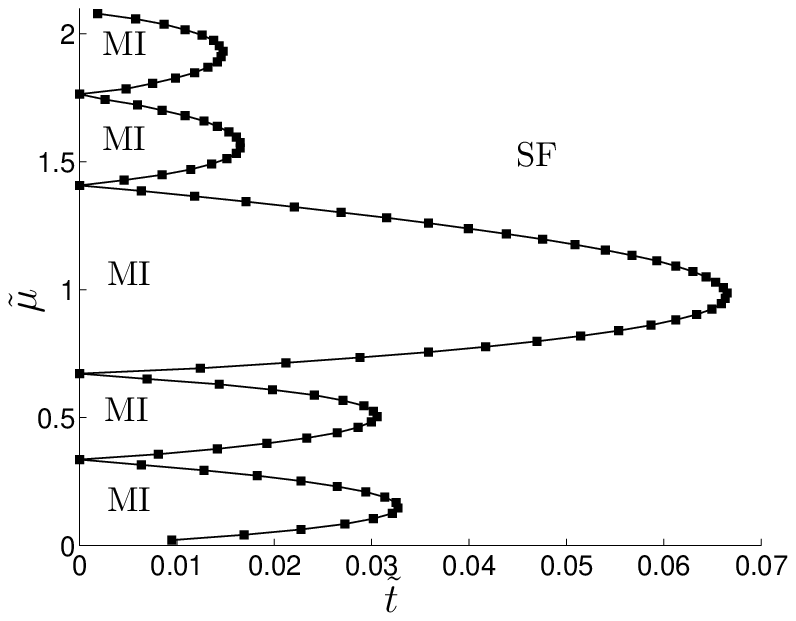}
\caption{
Mean field ground state phase diagram for the 3D
(three flavors) Hamiltonian.\label{fig:phasediag3D}}
\end{figure}

\subsection{Phase diagram 3D, three flavors}
Using the same type of mean field theory as for the 2D
(two flavors) case, the
3D (three flavors)
case can be treated as well. The resulting phase diagram for
$\tilde{U}_{xy}=1/3$ is shown in Fig.~\ref{fig:phasediag3D}.
Again, Mott-lobes with integer filling factors are seen surrounded
by a superfluid phase where all order parameters $\psi_{x,y,z}$
have equal magnitude, \emph{i.e.}, $|\psi_x|=|\psi_y|=|\psi_z|\neq 0$.
Due to the positivity of the coefficient $U_{xy}$ in the Josephson
term in Eq.~(\ref{eq:3dham}) the relative phases of the three
condensates are frustrated. Thus, writing
$\psi_{s}=|\psi_s|e^{i\phi_s}$, $s=x,y,z$ one finds
$\phi_x-\phi_y=\phi_y-\phi_z=\phi_z-\phi_x=\pm2\pi/3\pm \pi$.

 An interesting effect here is that the on site frustrated phase
configurations come in two different `chiralities' that cannot be
converted into each other by shifting any one of the phases by the
$\pi$ shift allowed by the $Z_2$ gauge symmetry. To see this one may
consider the current flowing between the condensates of different
flavors on a given site. The current flowing between the $X$ and $Y$
condensates on a particular site is determined by
$\sin(2[\phi_x-\phi_y])$. In a right handed configuration with, say,
$\phi_x=0,\phi_y=2\pi/3,\phi_z=4\pi/3$ there is an on-site current
flowing from
$$X\rightarrow Y\rightarrow Z\rightarrow X.$$
The situation is different in a left handed configuration with
$\phi_x=0,\phi_y=4\pi/3,\phi_z=2\pi/3$, where the current is now flowing in the opposite direction, \emph{i.e.}
$$X\leftarrow Y\leftarrow Z\leftarrow X.$$ Adding an arbitrary phase of $\pi$ (\emph{i.e.} invoking the $Z_2$ symmetry)
to any of the phases does not affect these currents.

Starting from a ground state with the same chirality throughout the
system one can choose a set of parallel planes and change the
chirality of each plane individually. Such changing of chirality of
a plane requires that the whole plane has the same chirality. This
additional ground state degeneracy is not associated with any
symmetry of the Hamiltonian but is an accidental one. A similar
situation occurs for special parameter values in frustrated
XY-models, where parallel zero energy domain walls can be
inserted~\cite{Korshunov}. One should note that such accidental
degeneracies at the mean field level may be lifted by fluctuation
effects associated with
collective modes such as spin waves.

As in the 2D (two flavors)
case, the smaller Mott-lobes, corresponding to integer
filling factors not divisible by the dimensionality of the system,
are degenerate in the $\tilde{t}=0$ limit. This degeneracy is lifted
due to tunneling, leading to (pseudo) magnetic ordering like that
demonstrated for the 2D
(two flavors)
case. To fully lift the degeneracy one has
to employ fourth order perturbation theory. The resulting
Hamiltonian will include terms acting simultaneously on three and
four sites. However, such fourth order corrections are very small
and may be difficult to observe in the proposed experimental
situation. They can however lead to novel physics and can be
intentionally generated~\cite{FisherZoller,coldatomToolBox}.

\begin{figure}[t]
\includegraphics[width=8cm]{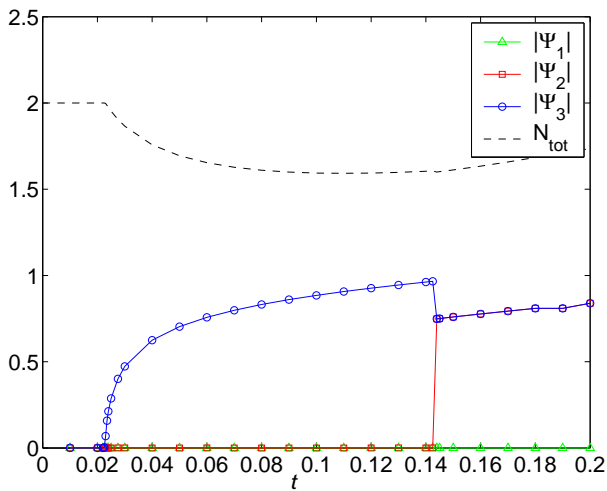}
\caption{
(Color online) Example of broken permutational symmetry. If one
increases the interspecies interaction $\tilde{U}_{xy}$ beyond
$1/3$, superfluid phases with broken permutational symmetry can be
achieved. Shown here are the order parameters sorted according to
magnitude, $\Psi_1=max(\psi_x,\psi_y,\psi_z)$,
$\Psi_3=min(\psi_x,\psi_y,\psi_z)$ for an interspecies interaction
$\tilde{U}_{xy}=0.8$ as a function of $\tilde{t}$. The total number
of particles is shown as a dashed line. \label{fig:cubsym}}
\end{figure}

Before leaving this section, we comment on the possibility of
breaking the permutational symmetry among the flavors in the
superfluid phase. As is well known, large interspecies interaction
strength in the two flavor bosonic Hubbard model leads to phase
separation. A phenomena occurring also here if $\tilde{U}_{xy}\ge
0.5$. However, due to the positive constant in front of the
'Josephson' (flavor changing) term, another phenomenon can take
place in the 3D (three flavors)
model.

As an example consider Fig.~\ref{fig:cubsym}. Here
$\tilde{\mu}=0.27$ and $\tilde{U}_{xy}=0.8$. As can be seen for
small $\tilde{t}$ the system is in a Mott insulating state with
filling factor 2. As $t$ increases the system becomes superfluid.
This occurs in two steps. First, mean field theory predicts a
second order transition to a state with only one flavor superfluid
and then a first order transition to a state with two nonzero
superfluid order parameters of equal magnitude. Increasing the
hopping strength further does not seem to make the third flavor
superfluid. We attribute this to the large energy cost associated
with having the phases of the three order parameters in a
frustrated configuration.

\section{Interference patterns and density correlations}
\label{sec:interference}
The traditional way of detecting superfluidity is by releasing the
trap and looking at the density distribution of the expanding
cloud.  Provided that the cloud expands many times its initial
diameter, the final position of a particle is determined by its
momentum rather than its initial position.  Hence this expanded
real-space density distribution provides a direct picture of the
momentum-space distribution of the trapped system. More precisely, the density distribution
a time $t$ after trap release is related to the momentum density of the trapped state $\left|\Phi\right>$
as
$$\left<n({\bf r},t)\right>=\left(\frac{m}{ht}\right)^3\left<\Phi|n_{\bf Q(r)}|\Phi\right>$$
where ${\bf Q(r)}=m{\bf r}/(\hbar t)$.
It is useful to think of this spatial distribution as resulting
from interference of matter waves radiated by the different
lattice sites when the trap is released.  The one-dimensional
character of the $Z_2$ gauge symmetry means that thermal
fluctuations can destroy the long range order phase order by
allowing the phase on an arbitrary site to flip by $\pm\pi$. If
the system disorders in this way, any interference pattern in the
radiated matter waves will be destroyed as well.
In this case, further information about the correlations in the
system can be obtained by looking at the density fluctuations
(noise) in the released
cloud~\cite{Altman:2004,Greiner:2005,Folling:2005} in a
Hanbury-Brown Twiss like statistical measurement.

We begin this section by looking at the zero temperature momentum
distribution and then consider the density fluctuations of the
expanded cloud around its mean.
\subsection{Interference patterns}
Although any real experiment is conducted at finite temperature,
the zero temperature columnar phase ordering may prevail for a
finite system at low enough temperatures. The zero temperature
momentum distribution is thus of interest and we will estimate it
by using a single macroscopically occupied wave function
corresponding to the superfluid states in the two- and three-flavor cases. The
details of the calculations can be found in the Appendix and we
here only state the main results.

\begin{figure}[t]
\includegraphics[width=8.0cm]{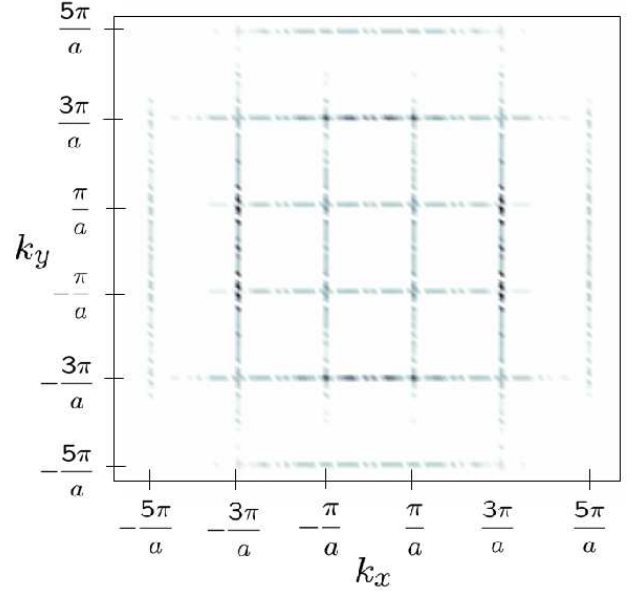} \caption{
Calculated momentum distribution for a 40x40 lattice. The momentum
distribution was calculated by numerically summing the contributions
to the distribution function.
\label{fig:momdist2d}}
\end{figure}

We begin by considering a single 2D plane with $N\times N$ sites at zero temperature in the two-flavor system
and model the superfluid state with a macroscopically occupied wavefunction
$$\left|\Phi\right>=\frac{\left(a_{SF}^\dagger\right)^M}{\sqrt{M!}}\left|0\right>.$$
Here $\left|0\right>$ is the vacuum state of the lattice, \emph{i.e.} no atoms present,
while $a_{SF}^\dagger$ is
the creation operator
$$a_{SF}^\dagger\equiv \frac{1}{\sqrt{2}N}\sum_{m=1}^N\sum_{n=1}^N
\left(\alpha_{mn}X_{mn}^\dagger+\beta_{mn}Y_{mn}^\dagger\right).$$
The subscripts $m$ and $n$ denote the coordinates, rows and
columns, in the lattice while $\alpha$ and $\beta$ are phase
factors ($|\alpha|=|\beta|=1$) determining the phase of the
wavefunction on a given site. At zero temperature the phases of
$X$- particles are ordered along rows while the phases of
$Y$-particles are ordered along columns, \emph{i.e.},
$\alpha_{mn}=\alpha_m$  and $\beta_{mn}=\beta_n$
(cf.~Fig.~\ref{fig:2dphases} and Fig.~\ref{fig:fields}). For a
macroscopic occupation $M$ the observed density distribution in a
single shot in the $x-y$ plane after expansion is proportional to
the momentum distribution $\left<\Phi\right|\psi_{\bf
Q}^\dagger\psi_{\bf Q}\left|\Phi\right>.$ As shown in the appendix
we have for a single 2D plane in the two-flavor system
$$\left<\Phi\right|\psi_{\bf Q}^\dagger\psi_{\bf Q}\left|\Phi\right>=|\tilde{\Psi}_x({\bf Q})|^2+|\tilde{\Psi}_y({\bf Q})|^2$$
where
\begin{eqnarray}
\left|\tilde{\Psi}_x\right|^2&=&\pi M\left|\tilde{\phi}_0^x({\bf Q})\right|^2
{f_1(Q_y,\alpha_m)}\sum_{{\rm odd}\,n}\delta\left(aQ_x-{n\pi}\right)\nonumber\\
\left|\tilde{\Psi}_y\right|^2&=&\pi M\left|\tilde{\phi}_0^y({\bf Q})\right|^2
{f_1(Q_x,\beta_n)}\sum_{{\rm odd}\,m}\delta\left(aQ_y-{m\pi}\right)\nonumber.
\end{eqnarray}
The functions $\tilde{\phi}_0^x$ and $\tilde{\phi}_0^y$ are the
Fourier transforms of the on-site Wannier functions and
${f_1(Q_y,\alpha_m)}$ and ${f_1(Q_x,\beta_m)}$ are
$2\pi/a$-periodic random functions with typical magnitude of order
unity which depend on which of the degenerate ground states is
observed (see Fig.~\ref{fig:random} and the Appendix for details).
From the above equations the interference pattern from a single 2D
plane in the two-flavor system can bee seen to be a grid like structure as shown in
Fig.~\ref{fig:momdist2d} where the interference pattern has been
calculated numerically for a 40x40 lattice. The appearance of
lines, rather than points as in a single flavor 2D system, stems
from the one-dimensional character of the superfluid state with
phases only being aligned along rows (columns) but randomly
distributed between rows (columns). The randomness in the
distribution between the rows (columns) show up as the random
interference pattern along grid lines.

In an experiment one typically does not probe a single plane but
it is the integrated density of a large number of planes that is
imaged. For imaging in the plane parallel to the 2D planes the
integrated column density (intensity in the absorption image) is
for a
$N\times N\times N$ lattice with $M$ atoms in each 2D plane
$$I(Q_x,Q_y)=N \int \frac{dQ_z}{2\pi} \overline{\left<\Phi\right|\psi_{\bf Q}^\dagger
\psi_{\bf Q}\left|\Phi\right>}.$$
Here the line over the quantum mechanical averaging denotes the averaging over
the different ground state configurations allowed by the $Z_2$ symmetry. Since
$\overline{f_1(Q,\alpha_m)}=1$ (see Appendix)
the random interferences seen in Fig.~\ref{fig:momdist2d} will be averaged out
and a grid of smooth lines, void of interference, will be seen. Another source
of smoothing out the random interferences comes from limited detector precision.
For a large system, the random oscillations becomes increasingly rapid and only an
average over nearby momenta can be probed.
\begin{figure}[t]
{\includegraphics[width=7cm]{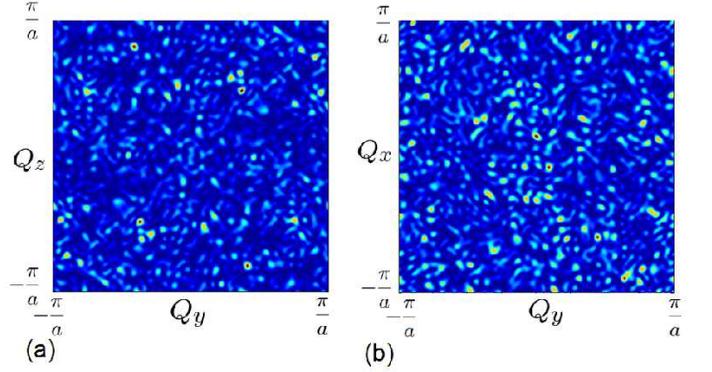} \caption{(color
online) (a) The random function $f_2(Q_y,Q_z,\eta_{no}^x)$ for a
$40\times 40$ lattice for a specific realization of $\eta_{no}^x$
Note that $f_2$ is symmetric under inversion. (b) The random
function $g_2^y(Q_x,Q_z,\eta_{no}^y,\sigma_m)$ for a $40\times 40$
lattice for a specific realization of $\eta_{no}^y$ and
$\sigma_m$. Note that $g_2$ is not symmetric under inversion.
\label{fig:f_2}}}
\end{figure}

In the 3D ({\em i.e.}
three flavors)
case, the situation is very
similar. Special care have to be taken with accidental symmetry
breaking of the ground state giving rise to planes of different
chirality. If we assume that planes with uniform chirality have
normals in the $x-$direction the momentum distribution can be
written
\begin{eqnarray}
&&\left<\Phi\right|\psi_{\bf Q}^\dagger\psi_{\bf Q}\left|\Phi\right>=
|\tilde{\Psi}_x({\bf Q})|^2+|\tilde{\Psi}_y({\bf Q})|^2+|\tilde{\Psi}_z({\bf Q})|^2\nonumber \\
&+&2{\rm Re}\left[\tilde{\Psi}_x({\bf Q})^*\tilde{\Psi}_y({\bf Q})\right]+2{\rm Re}
\left[\tilde{\Psi}_y({\bf Q})^*\tilde{\Psi}_z({\bf Q})\right]\nonumber\\
&+&2{\rm Re}\left[\tilde{\Psi}_z({\bf Q})^*\tilde{\Psi}_x({\bf Q})\right]
\label{eq:x}
\end{eqnarray}
where
\begin{eqnarray}
\left|\tilde{\Psi}_x\right|^2
&=&2\pi\left|\tilde{\phi}_0^x\right|^2{\frac{M}{3}}f_2(Q_y,Q_z,\eta_{no}^x)\sum_{{\rm odd}\, n}
\delta\left(aQ_x-{n}\pi\right)\nonumber
\end{eqnarray}
$$|\tilde{\Psi}_y|^2=2\pi|\tilde{\phi}_0^y|^2{\frac{M}{3}}g_2^y(Q_z,Q_x,\eta^y_m,\sigma_m)
\sum_{{\rm odd}\,n}\delta(aQ_y-n\pi)$$
$$|\tilde{\Psi}_z|^2=2\pi|\tilde{\phi}_0^z|^2{\frac{M}{3}}g_2^z(Q_x,Q_y,\eta^z_m,\sigma_m)
\sum_{{\rm odd}\,n}\delta(aQ_z-n\pi).$$ Again, since long range
order is only aligned along 1D strips, the released cloud will be
a set of intersecting perpendicular planes with intersections at
positions corresponding to odd momenta $Q_{x,y,z}=(2n+1)\pi/a$.
The planes in each direction will have a random intensity
modulation specified by the random functions $f_2$, $g_2^y$ and
$g_2^z$ (see Appendix for details). Examples of these distribution
functions $f_2$ and $g_2$ are shown in Fig.~\ref{fig:f_2}. The
last three terms in Eq.~\ref{eq:x} randomly modulate  the
distribution along the intersections of the planes.

If a single shot measurement is made, the integrated column
density will show a pattern of grid lines similar to that in
Fig.~{\ref{fig:momdist2d}}, the grid lines showing random
interference patterns. Between the lines a periodic random
distribution (of lesser intensity than the lines) will be present.
This latter distribution will be either $f_2$ or $g_2$ depending
on the orientation of the planes with uniform chirality. Thus if
the absorbtion image is taken in the same plane as the planes with
uniform chirality this background modulation will be symmetric
under space inversion (cf.~Fig~{\ref{fig:f_2}}(a)) whereas if it
is taken perpendicular there will be no such symmetry in the
random modulation (cf.~Fig~{\ref{fig:f_2}}(b)).

\subsection{Density-density correlations}
As pointed out above, the dimensional reduction present in the
system means that finite temperatures can destroy the 1D
Ising-like ordering of phases along columns and that the
individual phases at any one site can be flipped $\pm \pi$, i.e.,
the $Z_2$ gauge symmetry is restored. In this case there will be
no visible interference pattern although atoms are delocalized, \emph{i.e.}
the delta peaks will be smeared and a random density distribution
will be seen each shot.

To illustrate the usefulness of correlation measurements we
consider a single $N\times N$ 2D plane in the two-flavor system at unit filling ($M=N^2$).
If the temperature is finite, not only may the $Z_2$ symmetry in the superfluid state
be restored but it is also possible for the unit filling Mott state to be disordered.
There are then four different possible states the system can be in
\begin{enumerate}
\item Superfluid with restored $Z_2$ symmetry.
\item Ferromagnetic Mott insulator (all atoms of the same flavor).
\item Anti-ferromagnetic Mott insulator (alternating flavors on alternating sites).
\item Disorder Mott insulator (each site having one atom but with random flavor).
\end{enumerate}
If one makes multiple single shot measurements and averages the
density distribution obtained in each shot, one obtains a measure
of the average momentum distribution (see Appendix)
\begin{eqnarray}
\overline{\left<\Phi|n_{\bf Q}|\Phi\right>}&=&\frac{M}{2}\left[|{\tilde{\phi}_0^x}({\bf Q})|^2+
|{\tilde{\phi}_0^y}({\bf Q})|^2\right]\nonumber
\end{eqnarray}
which is the same for each of the four states 1-4.
We will henceforth refer to averages $\overline{\left<\cdot\right>}$ as \emph{disorder averages}.
To distinguish the four states
one can instead measure the HBT-like
density-density correlations of the expanding
cloud~\cite{Altman:2004,Greiner:2005,Folling:2005},
$$G({\bf r},{\bf r^\prime})\equiv \overline{\left<n({\bf r})n({\bf r^\prime})\right>}_t-\overline{\left<n({\bf r})
\right>}_t\overline{\left<n({\bf r^\prime})\right>}_t.$$
Here $\overline{\left<n({\bf r})\right>}_t$ is the density of atoms at
point ${\bf r}$ a time $t$ after the trap has been switched off averaged over many
experimental realizations (see Appendix). To measure $\overline{\left<n({\bf r})n({\bf r^\prime})\right>}_t$
 one calculates the product of the observed densities $n({\bf r})n({\bf r^\prime})$ in each shot and averages
 over several experimental runs.
Just as for the density distribution, the correlation function $G$
provides a measure of the momentum correlations
$$G({\bf r},{\bf r}')=\left(\frac{m}{ht}\right)^6\left[\overline{\left<n_{\bf Q}n_{\bf Q'} \right>}-
\overline{\left<n_{\bf Q}\right>}\times\overline{\left<n_{\bf Q'}\right>}\right]$$
prior to trap release.

To get a qualitative understanding of how the superfluid state can
be detected by correlation measurements we first return to the
$T=0$ result in the previous section and look at the periodic
function $f_1(Q_y)$. This random modulation arose because phases
of $Y$:s were uncorrelated between rows. Since the relative phases
of the $Y$:s in rows is $\pm \pi$ this function is even in $Q$ and
along any given grid line the quantity $\left<n_{\bf Q}n_{\bf
Q'}\right>$ is thus strongly correlated when $Q_y+Q_y'=2m\pi/a$.
Averaging over many realizations one sees that (along a grid line
of constant $Q_x=2n\pi/a$)
$$\overline{\left<n_{\bf Q}n_{\bf Q'}\right>}\Big|_{T=0}\propto \sum_{n}\delta(Q_y+Q_y'-2m\pi/a)+{\rm other\, terms}.$$
In the thermally disordered superfluid state the phase on any site is allowed to flip by $\pi$ (restoring $Z_2$ symmetry).
This destroys the delta peaks in $|\tilde{\Psi}_{(x,y)}|^2$ and gives instead a random modulation given by the function
$f_2(Q_x,Q_y)$. Since only $\pi$-flips are allowed this modulation is symmetric $f_2(Q_x,Q_y)=f_2(-Q_x,-Q_y)$ and we
get strong correlations in $\left<n_{\bf Q}n_{\bf Q'}\right>$
$$\overline{\left<n_{\bf Q}n_{\bf Q'}\right>}\Big|_{T\neq 0}\propto \sum_{n}\delta({\bf Q+Q'}-{\bf G}_n)+{\rm other\, terms}.$$
where ${\bf G}_n$ is a reciprocal lattice vector.
On a technical level (see Appendix) this can be seen to arise since the disorder averaged propagator for single
particles is necessarily
short-ranged due to the random $\pi$ phase changes at finite
temperature, while the disorder averaged propagators for {\em pairs} of particles can still
be long-ranged.

In Appendix we have calculated $G({\bf Q(r)},{\bf Q'(r')})$ for the four scenarios for a single plane
in the two-flavor system
and we here
give the qualitative results. The superfluid state is, confirming the qualitative discussion above, characterized by peaks
at ${\bf Q\pm Q'=G}_n$ (see Eq.~(\ref{eq:GSF2D})) while
the ferromagnetic Mott state and the disordered Mott state has peaks only at ${\bf Q- Q'=G}_n$ (see Eqs.~(\ref{eq:GFM2D})
and (\ref{eq:GDO2D})). The correlation function in these Mott states
are distinguished by having different background intensities between peaks and different peak strengths.
The antiferromagnetic Mott state has peaks at half reciprocal lattice vectors  ${\bf Q- Q'=G}_n/2$ (see Eq.~(\ref{eq:GAFM2D})).

In the three flavor system the situation is similar to the two flavor scenario discussed above. In the presence
of thermal disordering of the superfluid state the interference patterns of the Mott state and the superfluid states
become indistinguishable. Again, in the three flavor case the pair propagators will be nonzero in the disordered superfluid state and
 $G({\bf Q(r)},{\bf Q'(r')})$ will have peaks at ${\bf Q\pm Q'=G}_n$ (see Eq.~(\ref{eq:GDO3D}))

While there should be no problem to measure the correlation functions for a
system with three flavors,
the two-flavor system
poses a problem of
technical nature since in an experiment several uncorrelated 2D planes will be created. Suppose one has $N$ uncorrelated planes.
If one detects one atom at position ${\bf r}$ and another at ${\bf r}'$ in a single experiment the atoms could have come from
either the same plane or different planes. Measuring the
product of densities in each shot and averaging over several experiments one will for the case with $N$ 2D planes
actually measure
$$N\overline{\left<n({\bf r})n({\bf r'})\right>}_t+N(N-1)\overline{\left<n({\bf r})\right>}_t\overline{\left<n({\bf r'})\right>}_t$$
rather than $N\overline{\left<n({\bf r})n({\bf r'})\right>}_t.$ Thus the signal-to-noise ratio scales as $1/N$ requiring many experimental runs for large systems.

\section{Lifetime Estimate 1D\label{sec:lifetime}}
 In the previous sections, effective Hamiltonians for atoms in
the first band(s) of the optical lattice were introduced and the
mean field ground state phase diagrams drawn. In doing so, it was
assumed that the interaction terms in the original Hamiltonian
(\ref{eq:LatticeHamiltonian}) responsible for scattering particles
between bands could be ignored. In this section, these
interactions are taken into account perturbatively and the
lifetime of atoms in the first band is estimated. The obtained
(inverse) lifetime should be compared to other energy scales in
the problem, most importantly the smallest one, the hopping
energy. If the lifetime turns out to be long compared to the time
scale of hopping, the novel states described in the previous
sections should be possible to realize in experiment.

To simplify matters, the discussion will be restricted to the
1D-case. The ensuing results are expected to agree well, both
qualitatively as well as quantitatively, with the 2D and 3D cases to
lowest order in perturbation theory. This follows from taking parity
considerations into account when determining the allowed
transitions. Thus, ignoring tunneling in the $y-z$ directions and
measuring distance in units of the lattice spacing (see
section~\ref{sec:Lattice_Ham}) the 1D Hamiltonian can be written
$\hat{H}=\hat{H}_0+\hat{V}$ with
$$\hat{H}_0=E_R\sum_n \int
d\xi\,
 \hat{\psi}_{n}^\dagger(\xi)\left(-\frac{\partial^2}{\partial
{\xi}^2}+v_{0x}\sin^2(\xi)\right)\hat{\psi}_n(\xi)$$ and
\begin{equation}
\hat{V}=\frac{U}{2}\sum\limits_{n_1,n_2,n_3,n_4}\int d\xi\,
\hat{\psi}_{n_1}^\dagger(\xi)\hat{\psi}_{n_2}^\dagger(\xi)\hat{\psi}_{n_3}(\xi)\hat{\psi}_{n_4}(\xi).
\label{eq:1d}\end{equation} Here $\hat{\psi}_n^\dagger(\xi)$
creates an atom at $\xi$ in the \mbox{$n$:th} band of the
1D-system and $U \equiv 4\pi E_R
\left(\frac{a_s}{a}\right)O_{00}(v_{0y})O_{00}(v_{0z})$.

Apart from the field operators $\hat{\psi}_n(\xi)$ it is
convenient to define boson operators in two other bases. First, we
have the basis of Bloch functions $u_{nk}(\xi)$ with band index
$n$ and lattice-momentum $k$. These functions satisfy
$\hat{H}_0u_{nk}(\xi)=\epsilon_n(k)u_{nk}(\xi)$ and are associated
with the field operators $\hat{a}_{nk}, \hat{a}^\dagger_{nk}$.
Second, we have the Wannier-functions $\tilde{\phi}_n(\xi-\pi m)$
defined in section~\ref{sec:Lattice_Ham}. In this section we will denote the
corresponding field operators by
$\hat{a}_n(m),\hat{a}_n^\dagger(m)$. Note that these definitions
depart from the conventions in previous sections and that
operators corresponding to Bloch functions and Wannier functions
are distinguished in the number of subscripts.
\subsection{Wide-band limit}
Begin by looking at the case when the second term $\hat{V}$ in
eq.~(\ref{eq:1d}) is small compared to $\hat{H}_0$
and consider an initial state where all $N$ atoms reside in the lowest
lying Bloch state of the first band,
$$\left|i\right>=(N!)^{-1/2}{\left(\hat a^\dagger_{n=1,k=\pi/a}\right)^N}\left|0\right>.$$
A first order decay process is then one where two atoms in the
first band collide, promoting one to the second band, the other to
the zeroth band, i.e. the final state is
$$\left|f\right>=\frac{\hat a_{n=2,k_2}^\dagger \hat a_{n=0,k_0}^\dagger \hat a_{n=1,k=\pi/a}
\hat a_{n=1,k=\pi/a}}{\sqrt{N(N-1)}}\left|i\right>.$$ The first order
matrix element for this transition is
\begin{figure}[ht]
\includegraphics[width=8.0cm,clip]{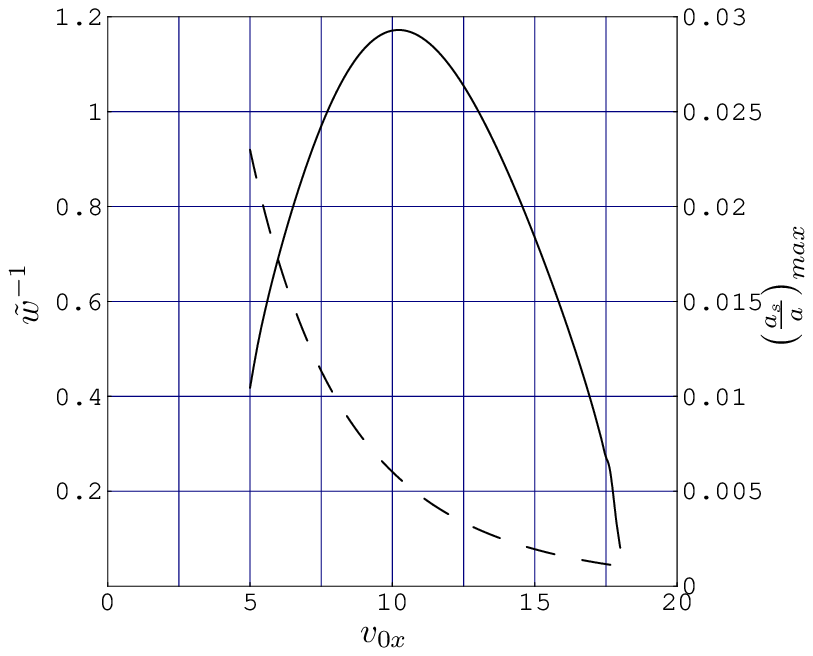} \caption{
(Solid line) Inverse of the coefficient $\tilde w$ occurring in
Eq.~(\ref{eq:decay}) for the first order decay rate calculated in
the wide-band limit. In this limit, there are no available energy
states for the first-order decay provided $19<v_{0x}$.(Dashed line)
The dashed line shows the maximum value of the ratio $a_s/a$ for
which  the wide-band analysis is valid.
\label{fig:lifetime}}
\end{figure}
\begin{eqnarray}
|\left<f\right|\hat{V}\left|i\right>|&=&U
\delta(k_0+k_2-2m\pi/a)\sqrt{N(N-1)}\nonumber \\
&\times&\int u_{0k_0}^*(\xi)u_{2k_2}^*(\xi)u_{1k=\pi/a}(\xi)^2
d\xi.\nonumber
\end{eqnarray}
If the filling factor (atoms/well) of the first band is $\nu_1$,
and the density of states of the $n$:th band is
$\rho(\epsilon_n(k))$ the transition rate per well becomes
$$w\approx\frac{2\pi}{\hbar}\frac{(\nu_1 U)^2 \left|\int u_{0k_0}^*(\xi)u_{2k_2}^*(\xi)u_{1k=\pi/a}(\xi)^2 d\xi
\right|^2}{\rho(\epsilon_0(k_0))^{-1}+\rho(\epsilon_2(k_2))^{-1}}.$$
Defining
$$\tilde{w}(v_{0z})\equiv \frac{32\pi^3 E_R}{\rho(\epsilon_0)^{-1}+\rho(\epsilon_2)^{-1}}\left|\int u_{0k_0}^*u_{2k_2}^*u_{1k=\pi/a}^2 d\xi
\right|^2$$ this can be compactly written as
\begin{equation}
w=\frac{E_R}{\hbar}\nu_1^2\left(\frac{a_s}{a}\right)^2
O_{00}(v_{0y})^2O_{00}(v_{0z})^2\tilde{w}(v_{0x}).
\label{eq:decay}
\end{equation}
In Fig.~\ref{fig:lifetime} $\tilde w(v)$ obtained from numerical
calculation is shown. For convenience the inverse of $\tilde w$
has been plotted. As can be seen, the lifetime goes to zero for
small and large $v$. This is a result of the diverging density of
states at the band edges. Above $v\approx 19$ the first order
process is no longer energetically possible and higher order
perturbation theory has to be applied.

The validity of the wide-band calculation relies upon the assumption that
the inequality $\nu_1 U< t$ is satisfied. This condition can be
used to obtain an upper bound on the ratio $a_s/a$ by assuming the
lattice depth to be the same in all directions, i.e.
$v_{0x}=v_{0y}=v_{0z}=v_0$, which yields
$$\frac{a_s}{a} <
\left(\frac{a_s}{a}\right)_{max}\equiv\frac{t}{\nu_1
E_R}\frac{1}{(2\pi)^{3/2}O_{00}(v_0)^2O_{11}(v_0)}.$$ This
quantity is shown for filling factor $\nu_1=1$ as the dashed line
in Fig.~\ref{fig:lifetime}.

\subsection{Narrow band limit}
\null From the discussion above it is clear that for deep enough
potentials the validity of the wide-band analysis breaks down unless
$\nu_1(a_s/a)$ is extremely small. An alternative starting point is
when $\nu_1 U\gg t$ while the filling factors for the zeroth and the
second band are small, i.e., $\nu_0,\nu_2 \ll 1$. Keeping terms of
order $\nu_1 U$, the relevant unperturbed Hamiltonian to start from
is in this case one where tunneling events in the two lowest bands
are completely ignored whereas interactions between atoms is only
considered for atoms interacting with particles in the first band.
Hence, one finds,
\begin{eqnarray}
\hat{H}_0&=&\sum\limits_{n=0,1}\sum\limits_m E_n(m)\hat n_n(m)+U_{0x}\sum\limits_m \hat{n}_0(m)\hat{n}_1(m)\nonumber \\ &+&\frac{1}{2}U_{xx}\sum\limits_m\hat{n}_1(m)(\hat{n}_1(m)-1)\nonumber \\
&+&E_R\sum\limits_{n>1} \int d\xi\,
\hat{\psi}_{n}^\dagger(\xi)\left(-\frac{\partial^2}{\partial
{\xi}^2}+v_{0x}(\xi)\right)\hat{\psi}_n(\xi)\nonumber
\\
&+&2U\sum\limits_{n>1}\int d\xi\,
\hat{\psi}_{1}^\dagger(\xi)\hat{\psi}_{1}(\xi)\hat{\psi}_{n}^\dagger(\xi)\hat{\psi}_{n}(\xi).
\label{eq:unpert}
\end{eqnarray}
Here the number operators $\hat{n}_n(m)\equiv\hat{a}^\dagger_n(m)\hat{a}_n(m)$ have been introduced.

\begin{figure}[ht]
\includegraphics[width=8cm]{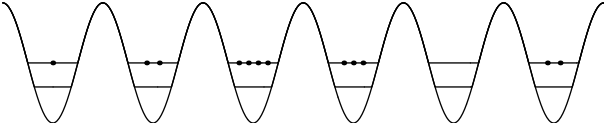}
\caption{
Typical initial state $\left|i\right>$ for the lifetime
estimate in the narrow band limit. All atoms are residing in the
first band, localized in the wells. In this particular case the
filling factors are $\nu_0=0$, $\nu_1=1$ and $\nu_{n>1}=0$.
\label{fig:inv}}
\end{figure}

The initial state is a product of Fock-states with definite
numbers of particles in the first band of each well as depicted in
Fig.~\ref{fig:inv}. Here, each well $m$ initially has $n_m$ atoms
in the first band, i.e.
$$\left|i\right>=\prod\limits_m \frac{\left(\hat
a^\dagger_1(m)\right)^{n_m}}{\sqrt{n_m!}}\left|0\right>.$$ The
final state is one where the population has changed such that, for
a particular well, denoted by $r$, one particle has decayed from
the first band down to the zeroth while another atom, in order to
conserve energy, ends up in a Bloch state of the $n$:th ($n>1$)
band, i.e.
$$\left|f\right>=\frac{\hat a_0^\dagger(r)\hat a^\dagger_{nk}\hat a_1(r)^2}{\sqrt{n_r(n_r-1)}}\left|i\right>.$$
This state is not an exact eigenstate of the unperturbed Hamiltonian
in Eq.~(\ref{eq:unpert}) but an approximate one. The correction to
the Bloch wave functions for $n>1$, which will later occur in the
overlap integrals, is however only of the order ${U}/{v_{0x}}\ll 1$
and can thus be ignored. What is more important is the associated
energy shift since this affects the position of the band edges of
the $n:th$ band. This, in turn, can have impact on the lifetime
since it affects the final density of states. Hence, one can replace
the two last terms in the unperturbed Hamiltonian~(\ref{eq:unpert})
by a term diagonal in the band index $n$;
\begin{eqnarray}\hat{H}_0&=&\sum\limits_{n=0,1}\sum\limits_m E_n(m)\hat n_n(m)+U_{0x}\sum\limits_m \hat{n}_0(m)\hat{n}_1(m)\nonumber \\ &+&\frac{1}{2}U_{xx}\sum\limits_m\hat{n}_1(m)(\hat{n}_1(m)-1)\nonumber \\
&+&\sum\limits_{n>1,k}(\epsilon_{n}(k)+\nu_1\Delta_{nk})\hat{n}_{nk}.
\end{eqnarray}
Here, the first order (Hartree) shift $\Delta_{nk}$ in energy due to interactions
between an atom in the $n:th$ Bloch band and the atoms in the
first band have been incorporated.
$$\Delta_{nk}\equiv U\int d\xi\, \left|u_{nk}(\xi)+u_{n,-k}(\xi)\right |^2\left|\tilde{\phi}_1(\xi)\right|^2$$

For first order decay one needs only the matrix element
$\left<f\right|\hat{V}\left|i\right>$ from which the rate follows;
\begin{equation}
w=\frac{2\pi}{\hbar}U^2n_r(n_r-1)\left|\int d\xi\,
u^*_{nk}(\xi)\tilde{\phi}^*_0(\xi)\tilde{\phi}_1(\xi)^2\right|^2\rho(\epsilon_n(k)).
\label{eq:narrowdecay}
\end{equation}

To present a comprehensive numerical analysis of this decay rate
is prohibitive due to the large number of parameters entering
expression Eq.~(\ref{eq:narrowdecay}). Thus, for sake of illustration,
we will here restrict the discussion to unit filling factor in the second band,
i.e. $\nu_1=1$. Further, we use
$(a_s/a)=1/100$ which is a reasonable value from an experimental
point of view. The lattice depths in the transverse directions
will be chosen slightly larger than in the $x$-direction, i.e.
chose $v_{0y}=v_{0z}=v_{0x}+1.$

\begin{figure}[t]
\includegraphics[width=7cm,clip]{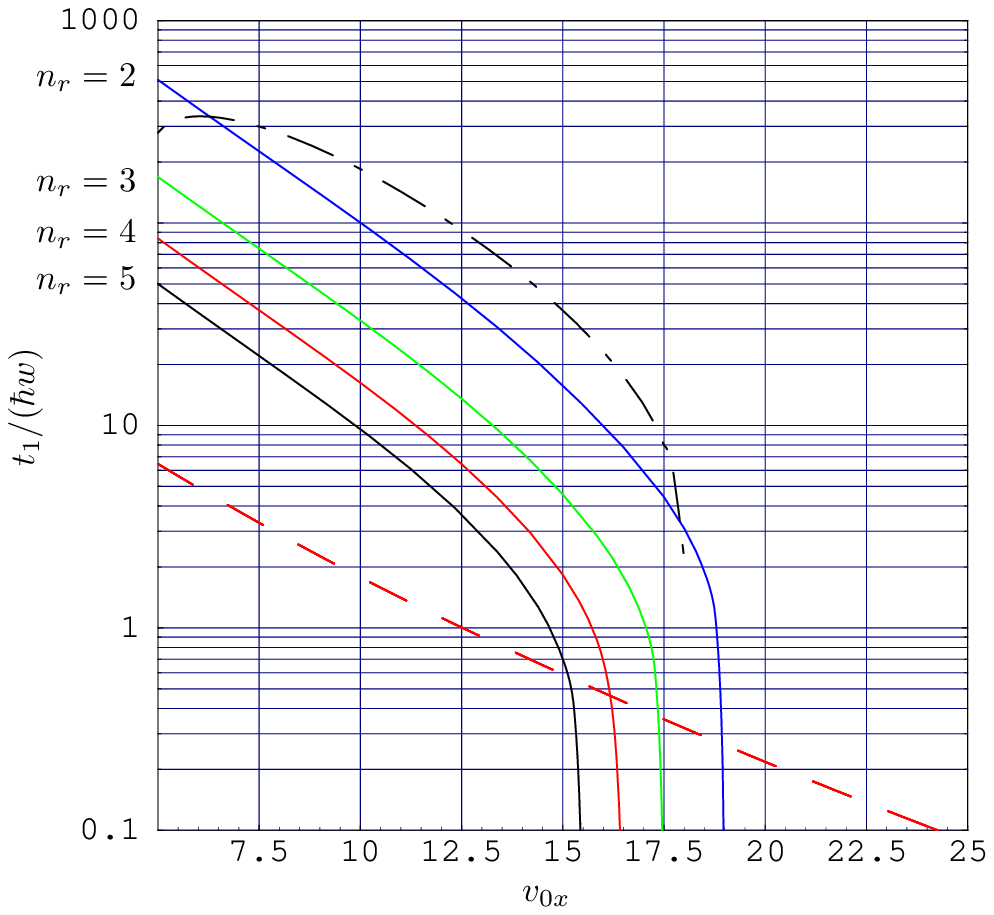}
\caption{
(Color online) First order lifetime $w^{-1}$ for a 1D system with
filling factor $\nu_1=1$ and $(a_s/a)=1/100$ in the narrow-band
limit according to Eq.~(\ref{eq:narrowdecay}). The solid lines show
the ratio between the lifetime and the time scale for hopping
($\hbar t_1^{-1}$) for $n_r$ particles in well $r$.
\null From top
to bottom; (blue) $n_r=2$; (green) $n_r=3$; (red) $n_r=4$; (black)
$n_r=5$. The dashed line shows the ratio $t_1/\nu_1 U$ which should
be less than unity for perturbation theory to be valid. The
dot-dashed line shows the result obtain by using the wide-band
formula in Eq.~(\ref{eq:decay}) using the same
parameters.\label{fig:decaynarrow}}
\end{figure}

The results of the calculation, using wave functions obtained from
band-structure calculations, are shown in Fig.~\ref{fig:decaynarrow}
which plots the ratio between the hopping rate and decay rate,
$t_1/(\hbar w)$. The different solid lines correspond to different
number of particles initially in the well. The cases $n_r=2,3,4,5$
are shown, $n_r=2$ having the longest lifetime and $n_r=5$ having
the shortest. The dashed line shows the ratio $t_1/(\nu_1U)$ which
should be less than unity for the expression to be valid. As a
comparison, the resulting lifetime obtained in the wide-band limit
Eq.~(\ref{eq:decay}) is also shown as the dash-dotted line.

The most interesting part of the result shown in
Fig.~(\ref{fig:decaynarrow}) is the sudden decay of the lifetime.
This is, as was the case in the wide-band limit, a result of the
diverging density of states $\rho(\epsilon_2(k))$ near the band
edge. For lattice potentials deeper than $v_{0x}\approx 20$, there
is no phase space (no available final energy levels for the
excited particle), available for the first order decay. To find
out the lifetime for larger values of $v_{0x}$, second order
perturbation theory is needed.

Consider again the same initial state $\left|i\right>$ as above.
Adhering to energy conservation arguments, there are three
different, mutually orthogonal, final states reachable through a
second order process,
$$\left|f_1\right>=\frac{\hat{a}_{nk}^\dagger\hat{a}_0^\dagger(r)^2\hat{a}_1(r)^3}
{\sqrt{2n_r(n_r-1)(n_r-2)}}\left|i\right>,$$
$$\left|f_2\right>=\frac{\hat{a}_{nk}^\dagger\hat{a}_0^\dagger(r)^3\hat{a}_1(r)^4}
{\sqrt{6 n_r(n_r-1)(n_r-2)(n_r-3)}}\left|i\right>,$$
$$\left|f_3\right>=\frac{\hat{a}_{n^\prime k^\prime}^\dagger\hat{a}_{nk}^\dagger\hat{a}_0^\dagger(r)^2\hat{a}_1(r)^4}
{\sqrt{2n_r(n_r-1)(n_r-2)(n_r-3)}}\left|i\right>.$$ The
corresponding decay rates $w_{1,2,3}$ are obtained from the
text-book relation
$$w_i=\frac{2\pi}{\hbar}\left|\sum\limits_{m}\frac{\left<f_i\right|V\left|m\right>\left<m\right|V\left|i\right>}
{\epsilon_i-\epsilon_m}\right|^2\rho(\epsilon_f).$$ Numerical
evaluations of the decay rates reveal that the dominant
contribution to the total decay rate $w_{tot}=w_1+w_2+w_3$ comes
from $w_1$. The reason for this is easily understood; the
contribution from $w_2$ is small due to destructive interference
of time reversed processes while the smallness of $w_3$, is due to
smallness of overlap integrals, which in turn can be understood
from parity considerations.

\begin{figure}[t]
\includegraphics[width=7.6cm]{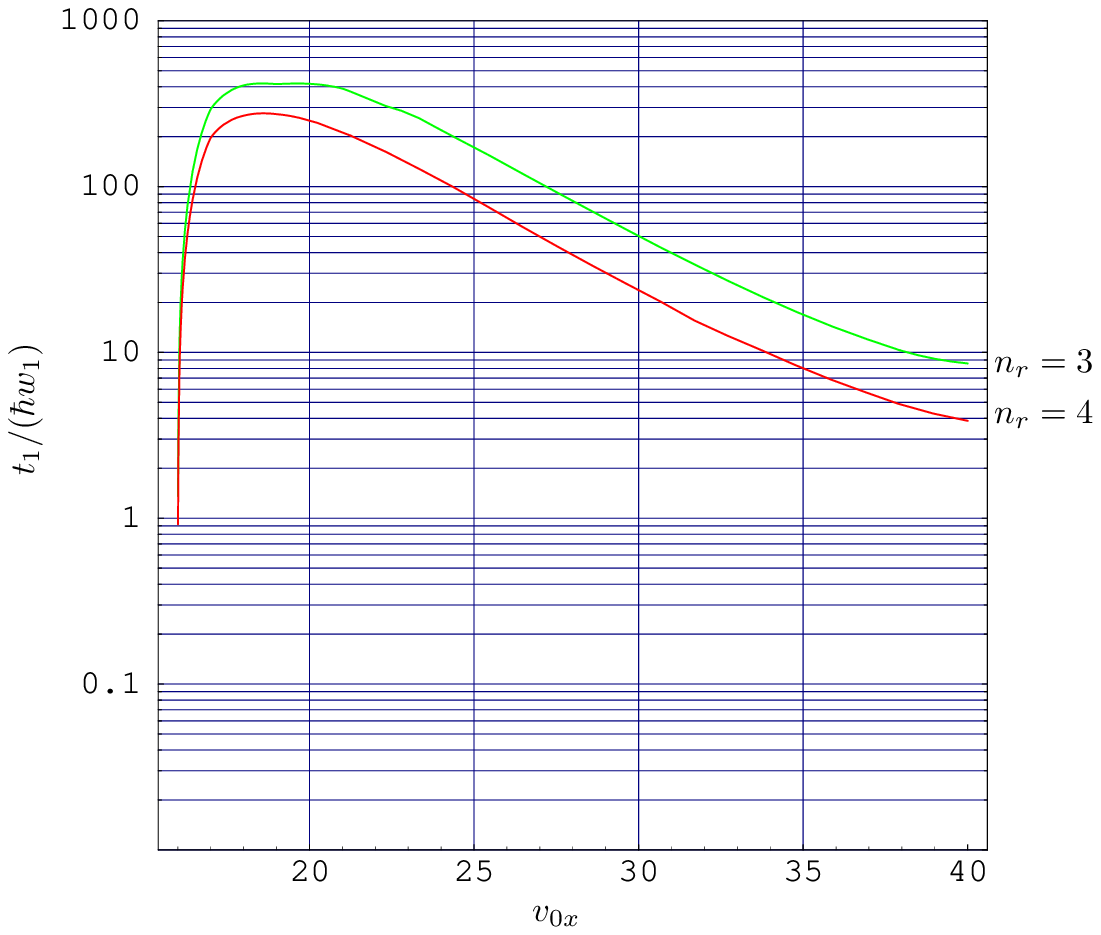}
\caption{
(Color online) Second order lifetime $w_1^{-1}$ for a 1D system with
filling factor $\nu_1=1$ and $(a_s/a)=1/100$ in the narrow-band
limit according to Eq.~(\ref{eq:narrowdecay}). The solid lines
show the ratio between the lifetime and the time scale for
hopping ($\hbar t_1^{-1}$) for $n_r$ particles in well $r$. From top to bottom; (green)
$n_r=3$; (red) $n_r=4$.  \label{fig:decaynarrow2}}
\end{figure}

The decay rate $w_1$ is shown in Fig.~\ref{fig:decaynarrow2}. As
can be seen, it is possible to achieve lifetimes considerably
larger than the inverse of the hopping energy, thus justifying
the validity of the Hamiltonians in
eqs.~(\ref{eq:2dham}) and (\ref{eq:3dham}).
\section{Conclusions}
By extending the usual mapping to the bosonic Hubbard model of
ultra cold atoms in an optical lattice to incorporate higher Bloch
bands, effective Hamiltonians governing the dynamics of atoms in
the first Bloch band(s) have been obtained. These Hamiltonians
resemble previously studied bosonic Hubbard Hamiltonian but differ
in two important respects:
\begin{itemize}
\item Atoms in the first
excited band are labelled by three possible flavors $X,Y,Z$.  The
dynamics is such that $X$ particles can (to a good approximation)
move only in the $x$ direction, etc.
\item Flavor changing collisions of atoms on the same site leading to
conversion of the form $XX\longrightarrow YY$, etc.~occur.
\end{itemize}
By appropriate choices of the lattice depths in the different directions the number
of flavors and the effective dimensionality (equal to the number
of flavors) of the system can be changed.
To obtain values of the relevant
parameters, such as hopping energy and interaction energies,
entering these effective Hamiltonians we have solved the time independent
Schr{\"o}dinger equation (Mathieu equation).

The effective Hamiltonians in two and three dimensions also show,
apart from the usual global $U(1)$ gauge symmetry,  a set of
$Z_2$-gauge symmetries intermediate between local and global. The
ground state in the 3D (three flavors) case also displays a chiral symmetry breaking
and an additional accidental ground state degeneracy associated with
different planar chiral ordering.

The phase diagrams for two particular cases relevant for experiment
have been sketched using mean field theory, indicating quantum
phase transitions between Mott-insulating and superfluid states.

Using time dependent perturbation theory up to second
order in the interatomic interactions the lifetime of the atoms in
the excited bands have been estimated. The results show that life
times considerably longer (orders of magnitude) than relevant
dynamical time scales can obtain. This suggests that it may be
possible to realize quasi-equilibrium in the subspaces of meta stable
states spanned by the effective Hamiltonians.
Finally, we would like to stress that the mean field theory used
to draw the phase diagram is only able to describe the most simple
scenario with a transition from a Mott-state to a superfluid state
with order parameter $\left<X\right>\neq 0$. It is well
known~[\onlinecite{Chen:2003,Kuklov:2002,Kuklov:2003,Altman:2003,Duan:2002}]
that other multi-flavor bosonic Hubbard models such as the
2-species Bose-Hubbard model shows a rich phase diagram with
phases which cannot be described in this simple approximation. The
present model, already rich at the mean-field level warrants
further study.  In particular, we have pointed out potential
connections to certain classes of models of frustrated
spins~\cite{Moore1,Moore2,Nussinov1,Nussinov2} and bose
metals~\cite{Paramekanti} that also have an infinite but
subextensive number of $Z_2$ gauge symmetries and as a result
exhibit dimensional reduction and exotic phases.
With the microscopic Hamiltonian
developed here, these connections can and should now be pursued in
detail.

\begin{acknowledgments}
The authors wish to acknowledge M.-C. Cha, K. Sengupta, N. Read
and S. Sachdev for many useful discussions. AI was in part
supported by The Swedish Foundation for International Cooperation
in Research and Higher Education (STINT) and SMG by the National Science
Foundation through NSF DMR-0342157.
\end{acknowledgments}

\appendix
\section*{Appendix}

Here we provide a detailed derivation of the density distribution one expects
to observe in the various phases and the different ways of measuring the density
density correlations in the released cloud of atoms.

If the system is in a many body quantum state $\left|\Phi\right>$ when the trap is released
at time $t=0$ the density distribution of atoms at a later time $t$ is given by
\begin{equation}
\left<n({\bf r})\right>_{t}=\left<\Phi\right|U^\dagger(t)n({\bf r})U(t)\left|\Phi\right>,
\label{eq:densdistr}
\end{equation}
where $U(t)$ is the time-evolution operator of the released system $U(t)=\exp({-i\hbar^{-1}Ht})$.
To measure the quantity in Eq.~(\ref{eq:densdistr}) one has to, in general, perform several measurements
starting with the same trapped state $\left|\Phi\right>$ each time. An exception to this is when the ground state
is to a good approximation a macroscopically occupied single particle state. This is typically the case for a superfluid system and a single measurement gives a good approximation of $\left<n({\bf r})\right>_{t}$.
For a weakly interacting dilute gas of atoms the interactions between atoms can be ignored during the expansion
of the cloud and the time evolution operator in Eq.~(\ref{eq:densdistr}) can be replaced by the free time-evolution
operator $U_0(t)$. Expanding in the momentum components one finds
\begin{eqnarray}
{\left<n({\bf r})\right>_t}&=&\int \frac{d{\bf k}_1}{(2\pi)^3}\int \frac{d{\bf k}_2}{(2\pi)^3} e^{-i({\bf k}_1-{\bf k}_2)\cdot ({\bf r}-\frac{\hbar t}{2m}({\bf k}_1+{\bf k}_2))}\nonumber\\
&\times&{\left<\Phi\right|\psi_{{\bf k}_1}^\dagger\psi_{{\bf k}_2}\left|\Phi\right>}.
\end{eqnarray}
For a system of linear size $L$ and for times $\hbar t\gg mL^2$ the stationary phase approximation gives
\begin{eqnarray}
{\left<n({\bf r})\right>_t}\approx\left(\frac{m}{h t}\right)^3\left<\Phi\right|n_{\bf Q({\bf r})}\left|\Phi\right>\quad {\bf Q}({\bf r})\equiv\frac{m{\bf r}}{\hbar t}.
\end{eqnarray}
Measuring the density of atoms after a long time of flight $t$ thus corresponds to a measurement of momentum
distribution of the state $\left|\Phi\right>$ prior to trap release.

In a typical experiment one takes an absorbtion image of the released cloud. This means that the only the {Integrated column density} is measured, \emph{ i.e.}, if an image of, say, the $x-y$ plane is taken, one measures
$$I(x,y)=\int dz {\left<n({\bf r})\right>_t}=\left(\frac{m}{h t}\right)^2\int \frac{dQ_z}{2\pi}\left<\Phi\right|n_{\bf Q({\bf r})}\left|\Phi\right>. $$
In the next subsection we derive the momentum distribution $\left<\Phi\right|n_{\bf Q({\bf r})}\left|\Phi\right>$
for the superfluid states in the two- and three-flavor systems
at zero temperature where $Z_2$ symmetry is broken.

\subsection{2D, two flavors, superfluid state, T=0}
For the two-flavor
case the system is comprised of 2D planes with uncorrelated ground states.
A superfluid state of a single 2D plane can be described by a wave function with $M$ particles in a single state
\begin{equation}
\left|\Phi\right>=({M!})^{-1/2}{\left(a_{SF}^\dagger\right)^M}\left|0\right>.
\label{eq:phistate}
\end{equation}
$$a_{SF}^\dagger\equiv \frac{1}{\sqrt{2}N}\sum_{m=1}^N\sum_{n=1}^N\left(\alpha_{mn}X_{mn}^\dagger+\beta_{mn}Y_{mn}^\dagger\right).$$
The subscripts $m$ and $n$ denote the coordinates, rows and columns, in the lattice while $\alpha$ and
$\beta$ are phase factors $|\alpha|=|\beta|=1$ determining the phase of the wavefunction on a given site.

\begin{figure}[t]
{\includegraphics[width=7cm,clip]{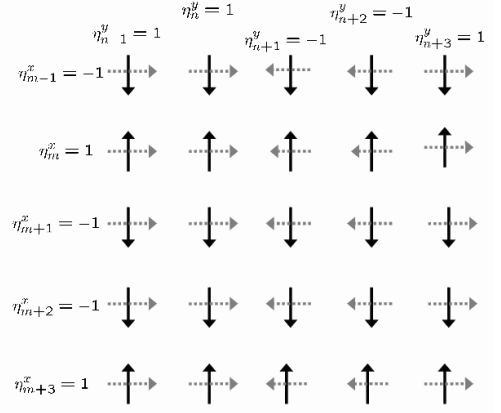}
\caption{Sample configuration of phases and the fields $\eta_m^x$ and $\eta_n^y$ for
a plane in the two-flavor
system at zero temperature.\label{fig:fields}}}
\end{figure}

\begin{figure}[t]
{\includegraphics[width=8cm,clip]{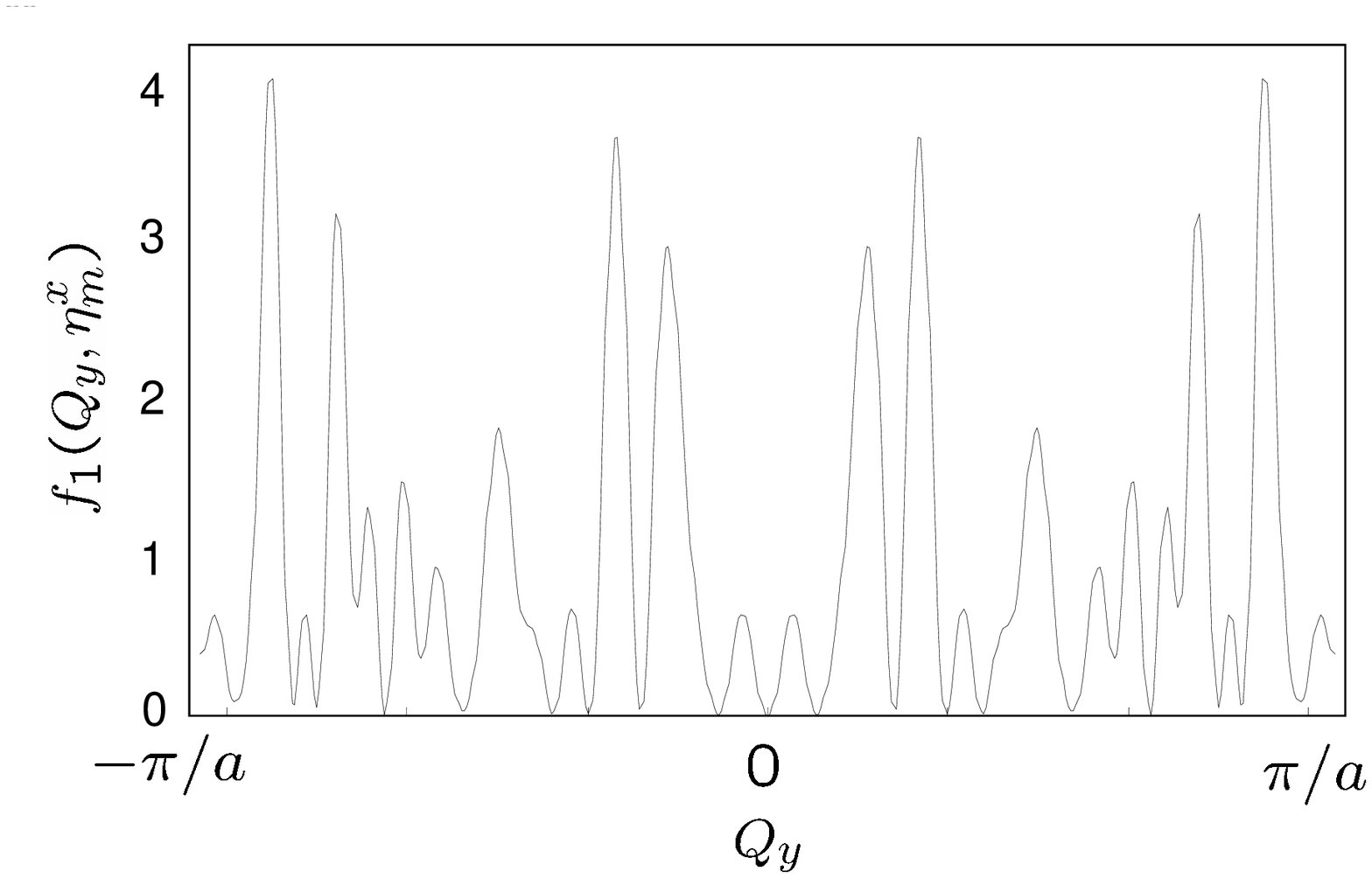}
\caption{Example of the random function $f_1(Q_y,\eta_m^x)$ defined in Eq.~(\ref{eq:doublesum}) .\label{fig:random}}}
\end{figure}

To evaluate the momentum distribution we expand the field operators $\psi_{\bf Q}^\dagger$ and $\psi_{\bf Q}$ in terms of the localized creation and destruction operators $X_{mn}^\dagger,Y_{mn}$ \emph{ etc.} where the subscripts
$m$ and $n$ respectively denote the row and column for the site on which the operator is acting.
For a general state $\left|\Phi\right>$ (not necessarily the state in Eq.~(\ref{eq:phistate})) we find
\begin{widetext}
\begin{eqnarray}
\left<\Phi\right|\psi_{\bf Q}^\dagger\psi_{\bf Q}\left|\Phi\right>&=&\int d{\bf r}_1d{\bf r}_2 e^{i{\bf Q}\cdot({\bf r}_1-{\bf r}_2)}\left<\Phi\right|\psi^\dagger({\bf r}_1)\psi({\bf r}_2)\left|\Phi\right>\nonumber=\sum_{m_1n_1}\sum_{m_2n_2}\int d{\bf r}_1\int d{\bf r}_2 e^{i{\bf Q}\cdot({\bf r}_1-{\bf r}_2)}\nonumber\\
&\times&\left<\Phi\right|\left[X^\dagger_{m_1n_1}{\phi_{m_1n_1}^x}({\bf r_1})^*+Y^\dagger_{m_1n_1}{\phi_{m_1n_1}^y}({\bf r_1})^*\right]\left[X_{m_2n_2}{\phi_{m_2n_2}^x}({\bf r_2})+Y_{m_2n_2}{\phi_{m_2n_2}^y}({\bf r_2})\right]\left|\Phi\right>.
\end{eqnarray}
The localized Wannier orbitals $\phi_{nm}^x({\bf r})$ and $\phi_{nm}^y({\bf r})$ can be rewritten
\begin{eqnarray}
\phi_{mn}^x({\bf r})&=&(-1)^n\phi_0^x({\bf r}-na\hat{x}-ma\hat{y}),\quad
\phi_{mn}^y({\bf r})=(-1)^m\phi_0^y({\bf r}-na\hat{x}-ma\hat{y})\nonumber
\end{eqnarray}
with the prefactors $(-1)^{n(m)}$ coming from the gauge choice in the inital way of writing the Hamiltonian in equation~\ref{eq:LatticeHamiltonian}. Carrying out the Fourier integrals we find
\begin{eqnarray}
\left<\Phi\right|\psi_{\bf Q}^\dagger\psi_{\bf Q}\left|\Phi\right>&=&\sum_{m_1n_1}\sum_{m_2n_2}
e^{i{\bf Q}\cdot({\bf R}_1-{\bf R}_2)}\left<\Phi\Big|\left[X^\dagger_{m_1n_1}(-1)^{n_1}{\tilde{\phi}_0^x}({\bf Q})^*+
Y^\dagger_{m_1n_1}(-1)^{m_1}{\tilde{\phi}_0^y}({\bf Q})^*\right]\right.\nonumber\\
&\times&\left.\left[X_{m_2n_2}(-1)^{n_2}{\tilde{\phi}_0^x}({\bf Q})+
Y_{m_2n_2}(-1)^{m_2}{\tilde{\phi}_0^y}({\bf Q})\right]\Big|\Phi\right>.
\label{eq:app1}
\end{eqnarray}
\end{widetext}
Here the position vectors ${\bf R}_1$ and ${\bf R}_2$ are shorthand for the lattice vectors
$${\bf R}_1\equiv n_1 a \hat{x}+m_1 a \hat{y}, \quad {\bf R}_2\equiv n_2 a \hat{x}+m_2 a \hat{y}$$
and $\tilde{\phi}_0^{x(y)}({\bf Q})$ denote the Fourier transform of the onsite wavefunctions.
In the Harmonic oscillator approximation these are given by
\begin{eqnarray}
\tilde{\phi}_0^x({\bf Q})&=& \pi^{3/4 }\gamma^{-5/2}Q_xe^{-\frac{Q_x^2+Q_y^2+Q_z^2}{2\gamma^2}},\quad
\gamma^2=\frac{2\pi\sqrt{mV_0}}{\hbar\lambda}\nonumber\\
\tilde{\phi}_0^y({\bf Q})&=& \pi^{3/4 }\gamma^{-5/2}Q_ye^{-\frac{Q_x^2+Q_y^2+Q_z^2}{2\gamma^2}}
\end{eqnarray}
To evaluate $\left<\Phi\right|\psi_{\bf Q}^\dagger\psi_{\bf Q}\left|\Phi\right>$ we need to calculate
the expectation values of the kind $\left<\Phi\right|X^\dagger Y\left|\Phi\right>$. For the superfluid state $\left|\Phi\right>$ in Eq.~(\ref{eq:phistate}) it is easily verified that in terms of the in-plane density $\rho\equiv M/N^2$ one gets
\begin{eqnarray}
\left<\Phi\Big|X^\dagger_{m_1n_1}X_{m_2n_2}\Big|\Phi\right>&=&\frac{\rho}{2}\alpha_{m_1n_1}^*\alpha_{m_2n_2}\nonumber\\ \left<\Phi\Big|X^\dagger_{m_1n_1}Y_{m_2n_2}\Big|\Phi\right>&=&\frac{\rho}{2}\alpha_{m_1n_1}^*\beta_{m_2n_2}\nonumber
\end{eqnarray}
\emph{etc.} Hence the terms in Eq.~(\ref{eq:app1}) factors and one can write it conveniently as
\begin{eqnarray}
\left<\Phi\right|\psi_{\bf Q}^\dagger\psi_{\bf Q}\left|\Phi\right>&=&
|\tilde{\Psi}_x({\bf Q})|^2+|\tilde{\Psi}_y({\bf Q})|^2\nonumber \\
&+&2{\rm Re}\left[\tilde{\Psi}_x({\bf Q})^*\tilde{\Psi}_y({\bf Q})\right]
\label{eq:app3}
\end{eqnarray}
where we have defined
\begin{eqnarray}
\tilde{\Psi}_x({\bf Q})&\equiv& \tilde{\phi}_0^x({\bf Q})\sqrt{\frac{\rho}{2}}\sum_{mn}e^{-i{\bf Q}\cdot {\bf R}_{mn}}(-1)^n\alpha_{mn}\label{eq:psirep1}\\
\tilde{\Psi}_y({\bf Q})&\equiv& \tilde{\phi}_0^y({\bf Q})\sqrt{\frac{\rho}{2}}\sum_{mn}e^{-i{\bf Q}\cdot {\bf R}_{mn}}(-1)^m\beta_{mn}\label{eq:psirep2}
\end{eqnarray}
For a system at absolute zero the phase factors $\alpha$ and $\beta$ are aligned along rows and columns
respectively but are randomly distributed between the lines and columns. To describe this situation we introduce two sets of fields, $\eta_m^x$ and $\eta_n^y$ which can take on values $\pm 1$. The relation between these values of the fields and the phases along rows and columns is shown in Fig.~\ref{fig:fields}.
Thus we can write
\begin{equation}
\alpha_{nm}=\eta_{m}^x\quad\beta_{nm}=i\eta_{n}^y.
\label{eq:phasefields}
\end{equation}
Consider now the summations needed to evaluate $\tilde{\Psi}_x$
\begin{eqnarray}
\tilde{\Psi}_x({\bf Q})=\tilde{\phi}_0^x({\bf Q})\sqrt{\frac{\rho}{2}}\sum_{mn}e^{-i{\bf Q}\cdot {\bf R}_{mn}}(-1)^n\eta_{m}^x
\end{eqnarray}
The summation over columns ($n$-summation) converges in the large $N$ limit to a sequence of delta-functions
\begin{eqnarray}
\tilde{\Psi}_x\equiv 2\pi\tilde{\phi}_0^x \sqrt{\frac{N\rho}{2}}\sum_{mn}\delta\left(aQ_x-{(2n+1)\pi}\right)e^{-i{mQ_ya}}\eta_{m}^x\nonumber\\
\label{eq:Psi_x}
\end{eqnarray}
and a similar equation can be obtained for $\tilde{\Psi}_y$.
Hence
\begin{eqnarray}
\left|\tilde{\Psi}_x\right|^2&=&2\pi N\frac{\rho}{2}\left|\tilde{\phi}_0^x\right|^2\sum_{n}\delta\left(aQ_x-{(2n+1)\pi}\right)\nonumber\\
&\times&\sum_{mm'}e^{-i{(m-m')Q_ya}}\eta_{m}^x\eta_{m'}^x.\nonumber
\end{eqnarray}
Introducing $\Delta=m-m'$ the last summations can be rewritten
\begin{eqnarray}
Nf_1(Q_y,\eta_m^x)&\equiv&\sum_{mm'}e^{-i{(m-m')Q_ya}}\eta_{m}^x\eta_{m'}^x\nonumber\\
&=&\sum_{\Delta}e^{-i{\Delta Q_ya}}\sum_{m}\eta_{m}^x\eta_{m-\Delta}^x\nonumber\\
&=&N+\sum_{\Delta\neq 0}e^{-i{\Delta Q_ya}}\sum_{m}\eta_{m}^x\eta_{m-\Delta}^x
\label{eq:doublesum}
\end{eqnarray}
where we have defined the random momentum distribution function $f_1$.
With the aid of Eq.~(\ref{eq:doublesum}) we now deduce some properties of $f_1$.
We begin with the magnitude of the function for any value of $Q_y$.
For each nonzero value of $\Delta$ the last summation is over an uncorrelated sequence of integers $\pm 1$ and can be viewed as a 1D random walk for which we have that
$$\sum_{m}e^{i\pi(\eta_{m}^x-\eta_{m-\Delta}^x)}\sim {\cal O}(\sqrt{N})$$
The summation over $\Delta$ contains $N-1$ terms which for each value of $Q_y$ are random of magnitude $\sqrt{N}$.
This is again a random walk with $N-1$ steps and we conclude that the whole expression in Eq.~(\ref{eq:doublesum}) is of order $N$. This can also be seen by noting that
\begin{eqnarray}
\frac{a}{2\pi}\int_{-\pi/a}^{\pi/a} dQ_y f_1(Q_y,\eta_m^x)=1.\nonumber
\end{eqnarray}
Thus for each configuration $\eta_m^x$ we have a randomly oscillating function $f(Q_y,\eta_m^x)$ of unit magnitude. An example
of $f_1$ obtained for a specific realization of $\eta_m^x$ with $N=40$ is shown in Fig.~\ref{fig:random}.
From Eq.~(\ref{eq:doublesum}) it is also clear, since $\eta_m^x$ and $\eta_{m-\Delta}$ are uncorrelated for nonzero $\Delta$ that the average over allowed ground state configurations is
$$\overline{f_1(Q_y,\eta_m^x)}\equiv \frac{1}{2^N}\sum_{\eta_1^x,\cdots,\eta_N^x=\pm 1}f_1(Q_y,\eta_m^x)=1.$$
An important property of $f_1$ is that it is even in $Q_y$
$$f_1(Q_y,\eta_m^x)=f_1(-Q_y,\eta_m^x).$$
This is a result of the nematic ordering between rows.
$|\tilde{\Psi}_y|^2$ can be calculated the same way as $|\tilde{\Psi}_x|^2$ and we get
\begin{eqnarray}
\left|\tilde{\Psi}_x\right|^2&=&\pi M\left|\tilde{\phi}_0^x({\bf Q})\right|^2{f_1(Q_y,\eta_m^x)}\sum_{{\rm odd}\, n}\delta\left(aQ_x-{n\pi}\right)\nonumber\\
\left|\tilde{\Psi}_y\right|^2&=&\pi M\left|\tilde{\phi}_0^y({\bf Q})\right|^2{f_1(Q_x,\eta_n^y)}\sum_{{\rm odd} \, m}\delta\left(aQ_y-{m}\pi\right)\nonumber
\end{eqnarray}
The interference term, the last part of Eq.~(\ref{eq:app3}), for the momentum distribution vanishes. To see this one can make use of equations (\ref{eq:phasefields}) and (\ref{eq:Psi_x})
\begin{eqnarray}
&&2{\rm Re}\left[\tilde{\Psi}_x({\bf Q})^*\tilde{\Psi}_y({\bf Q})\right]\nonumber\\
&=&4\pi^2\rho{\tilde{\phi}_0^x}\tilde{\phi}_0^y  {\rm Re}\left[ \sum_{mn}\delta\left(aQ_x-{(2n+1)\pi}\right)e^{i{mQ_ya}}\eta_{m}^x\right.\nonumber\\
&\times&\left. i\sum_{mn}\delta\left(aQ_y-{(2m+1)\pi}\right)e^{-i{nQ_xa}}\eta_{n}^y\right]=0\nonumber
\end{eqnarray}
\subsection{3D, three flavors, superfluid state, T=0}
For the three flavor
case at T=0 we consider again a state of the kind in Eq.~(\ref{eq:phistate}) but with
$$a_{SF}^\dagger\equiv \frac{1}{\sqrt{3}N^{3/2}}\sum_{j=1}^{N^3}\left(\alpha_{j}X_{j}^\dagger+\beta_{j}Y_{j}^\dagger+\gamma_{j}Z_{j}^\dagger\right).$$
The subscript $j$ denotes collectively the $x$- $y$- and $z$-coordinates in the 3D lattice. As in the
two flavor case, the observed momentum distribution can be written as
\begin{eqnarray}
\left<\Phi\right|\psi_{\bf Q}^\dagger\psi_{\bf Q}\left|\Phi\right>&=&
|\tilde{\Psi}_x({\bf Q})|^2+|\tilde{\Psi}_y({\bf Q})|^2+|\tilde{\Psi}_z({\bf Q})|^2\nonumber \\
&+&2{\rm Re}\left[\tilde{\Psi}_x({\bf Q})^*\tilde{\Psi}_y({\bf Q})\right]\nonumber\\
&+&2{\rm Re}\left[\tilde{\Psi}_y({\bf Q})^*\tilde{\Psi}_z({\bf Q})\right]\nonumber\\
&+&2{\rm Re}\left[\tilde{\Psi}_z({\bf Q})^*\tilde{\Psi}_x({\bf Q})\right]
\label{eq:app5}
\end{eqnarray}
with
$$\tilde{\Psi}_x({\bf Q})\equiv \tilde{\phi}_0^x({\bf Q})\sqrt{\frac{\rho}{3}}\sum_{mno}e^{-i{\bf Q}\cdot {\bf R}_{mno}}(-1)^m\alpha_{mno}$$
$$\tilde{\Psi}_y({\bf Q})\equiv \tilde{\phi}_0^y({\bf Q})\sqrt{\frac{\rho}{3}}\sum_{mno}e^{-i{\bf Q}\cdot {\bf R}_{mno}}(-1)^n\beta_{mno}$$
$$\tilde{\Psi}_z({\bf Q})\equiv \tilde{\phi}_0^z({\bf Q})\sqrt{\frac{\rho}{3}}\sum_{mno}e^{-i{\bf Q}\cdot {\bf R}_{mno}}(-1)^o\gamma_{mno}$$

Here the subscripts $mno$ refer to the $x-$ $y-$ and $z-$ coordinates in the lattice respectively.
To see how to handle the phase factors in the
three flavor
case, we begin with a state without accidentally broken chiral
symmetry
$$\alpha_{mno}=\eta_{no}^x, \beta_{mno}=e^{i2\pi/3}\eta_{mo}^y, \gamma_{mno}=e^{i4\pi/3}\eta_{mn}^z$$
where the random fields $\eta_{ij}$ can again take on values $\pm 1$.
Since the accidental chiral symmetry breaking occurs in parallel planes we can without loss of generality single
out the $x$-direction as the direction in which planes have uniform chirality (To compare with Fig~\ref{fig:3dphases} make the rotation of axes $y\rightarrow z, z\rightarrow x, x\rightarrow y$ in Fig.~\ref{fig:3dphases}). We thus introduce an additional
field $\sigma_m$ taking values $\pm 1$ for planes with different $x$-coordinate $m$. The corresponding phase factors for such a state will be
\begin{eqnarray}
\alpha_{mno}=\eta_{no}^x,\, \beta_{mno}=e^{i\sigma_m\frac{2\pi}{3}}\eta_{mo}^y,\,
\gamma_{mno}=e^{i\sigma_m \frac{4\pi}{3}}\eta_{mn}^z\label{eq:chiralfield}
\end{eqnarray}
We can now evaluate $\left|\tilde{\Psi}_x\right|^2$ in the same way as for the
two flavor case
\begin{eqnarray}
\left|\tilde{\Psi}_x\right|^2
&=&2\pi\left|\tilde{\phi}_0^x\right|^2{\frac{M}{3}}f_2(Q_y,Q_z,\eta_{no}^x)\sum_{{\rm odd}\, m}\delta\left(aQ_x-{m}\pi\right)\nonumber\\
\label{eq:2Dx}
\end{eqnarray}
In Eq.~(\ref{eq:2Dx}) $f_2(Q_y,Q_z,\eta_{no}^x)$ has been introduced
\begin{eqnarray}
f_2&\equiv&\frac{1}{N^2}\sum_{\stackrel{n_1o_1}{n_2o_2}}e^{ia(n_1-n_2)Q_y}e^{ia(o_1-o_2)Q_z}\eta_{n_1o_1}^x\eta_{n_2o_2}^x.\nonumber
\end{eqnarray}
The random distribution function $f_2$ is the two-variable analog of the function $f_1$ above. An example of $f_2$ for a 40x40 lattice is shown in Fig.~\ref{fig:f_2}(a). Just as $f_1$, $f_2$ obeys a sum rule
$$\left(\frac{a}{2\pi}\right)^2\int_{-\pi/a}^{\pi/a}\int_{-\pi/a}^{\pi/a} dQ_ydQ_zf_2(Q_y,Q_z,\eta_{no}^x)=1$$
is symmetric under inversion
$$f_2(Q_y,Q_z,\eta_{no}^x)=f_2(-Q_y,-Q_z,\eta_{no}^x)$$
and has an average equal to unity when averaged over ground states
$$\overline{f_2(Q_y,Q_z,\eta_{no}^x)}=1.$$
The expressions for $\left|\tilde{\Psi}_y\right|^2$ and $\left|\tilde{\Psi}_y\right|^2$ are similar but the
accidental ground state degeneracy modifies the random distribution functions. Explicitly we have
\begin{eqnarray}
|\tilde{\Psi}_y|^2&=&2\pi|\tilde{\phi}_0^y|^2{\frac{M}{3}}g_2^y(Q_z,Q_x,\eta^y_m,\sigma_m)\sum_{{\rm odd}\,n}\delta(aQ_y-n\pi)\nonumber\\
|\tilde{\Psi}_z|^2&=&2\pi|\tilde{\phi}_0^z|^2{\frac{M}{3}}g_2^z(Q_x,Q_y,\eta^z_m,\sigma_m)\sum_{{\rm odd}\,n}\delta(aQ_z-n\pi)\nonumber
\end{eqnarray}
with
\begin{eqnarray}
g_2^y&\equiv&\sum_{\stackrel{m1o_1}{m_2o_2}}e^{ia[(m_1-m_2)Q_x+(o_1-o_2)Q_z]}
\eta_{m_1o_1}^y\eta_{m_2o_2}^ye^{-i\frac{2\pi}{3}(\sigma_{m_1}-\sigma_{m_2})}\nonumber\\
g_2^z&\equiv&\sum_{\stackrel{m1n_1}{m_2n_2}}e^{ia[(m_1-m_2)Q_x+(n_1-n_2)Q_y]}
\eta_{m_1n_1}^z\eta_{m_2n_2}^ze^{-i\frac{4\pi}{3}(\sigma_{m_1}-\sigma_{m_2})}.\nonumber
\end{eqnarray}
En example of the distribution function $g_2^y$ is shown in Fig.~\ref{fig:f_2}(b). Note that
due to the fields $\sigma$ characterizing the different chirality of planes this distribution function
is \emph{not} symmetric under inversion.
Finally we look at the interference terms in Eq.~(\ref{eq:app5}).
\begin{widetext}
\begin{eqnarray}
\tilde{\Psi}_x^*\tilde{\Psi}_y
&=&4\pi^2\tilde{\phi}_0^x\tilde{\phi}_0^y\frac{\rho}{3}
\sum_{m,n\,{\rm odd}}\delta(aQ_x-m\pi)\delta(aQ_y-n\pi)\sum_{n_1o_1}\sum_{m_2o_2}e^{iaQ_z(o_1-o_2)}(-1)^{n_1}\eta_{n_1o_1}^x(-1)^{m_2}\eta_{m_2o_2}^ye^{i\frac{2\pi}{3}\sigma_{m_2}}\nonumber\\
\end{eqnarray}
\end{widetext}
In the above equation the summations over $n_1$ and $m_1$ constitute random walks. For the $n_1$ summation this is
a random walk on a line with $N$ unit steps $\pm 1$ giving rise to, for each value of $o_1$ a random term of order $\sqrt{N}$. The sum over $m_2$ can also be viewed as a random walk for each value of $o_2$ but in the complex plane. Each step being of unit length in any of the four directions $\pm 2\pi/3$ and $\pm 4\pi/3$. Summing over $n_1$ and $m_2$ thus yields, for each $(o_1,o_2)$ a random term of magnitude $N$ with a completely random phase.
Thus the interference terms in Eq.~(\ref{eq:app5}) (the other two terms can be treated similarly) give rise to a three dimensional grid of lines in the released cloud where the density along any given line is randomly distributed. If the density is averaged over several shots, with different ground states we have no contribution from the interference terms since
$\overline{\tilde{\Psi}_x^*\tilde{\Psi}_y}=0$.
\subsection{Density averages and correlations , $T\gtrsim 0$}
If $T$ is large enough for thermal fluctuations to restore $Z_2$ symmetry but still small enough to preserve the distinction between the Mott state and the superfluid state, measuring the density distribution alone does not suffice since the delta-peaks will be smeared.
Instead correlations can be measured. To this end, assume we have a single physical system. At finite $T$ this system undergoes transitions in a manifold of $\cal N$ states. Denote this manifold by the states $\left\{\left|\Phi_i\right>\right\}_{i=1}^{\cal N}$. In a single shot a single one of these states will be probed.
In an infinite series of experiments each of these states will be probed an infinite number of times and
one can thereby measure the quantity
\begin{equation}
\overline{\left<\Phi\right|\hat{O}\left|\Phi\right>}\equiv \frac{1}{\cal N}\sum_{i=1}^{\cal N}\left<\Phi_i\right|{\hat O}\left|\Phi_i\right>.
\label{eq:disorderaverage}
\end{equation}
Here we have ignored the Boltzmann factors since the manifold we are looking at is nearly degenerate.
In reality only a finite sequence of ${\cal M}$ experiments can be carried out and the fluctuations in $\overline{\left<\Phi\right|\hat{O}\left|\Phi\right>}$ is of concern. There are two sources fluctuations; First, for each state $\left|\Phi_i\right>$ there is quantum shot noise. Second, since not all of the ${\cal N}$ states will be probed there will be deviations due to not sampling the entire distribution.

If the manifold of states probed are superfluid states then ${\cal N}={\cal O}([2d]^{Nd})$ with $d$ being the dimensionality of the system. Since superfluid states are to a good approximation macroscopically occupied single particle states, fluctuations due to shot noise are reduced. The remaining fluctuations are classical and expected to scale as ${\cal M}^{-1/2}$ and
$\overline{\left<\Phi\right|\hat{O}\left|\Phi\right>}$ should in principle be possible to measure.
On the other hand, if the state measured is a Mott state the manifold $\left\{\left|\Phi_i\right>\right\}_{i=1}^{\cal N}$ consists typically of only a few states in which case multiple measurements reduces the quantum shot noise
since each quantum state will be probed many times.
We thus conclude, that by making repeated measurements and averaging the results, one can measure
$\overline{\left<\Phi\right|\hat{O}\left|\Phi\right>}$.

A quantity of interest to measure in this way is the correlation function
$$G({\bf r},{\bf r}')\equiv\overline{\left<n({\bf r})n({\bf r}')\right>}_t-\overline{\left<n({\bf r})\right>}_t\times\overline{\left<n({\bf r}')\right>}_t.$$
Again, if $\hbar t\gg mL^2$ this is to a good approximation the same as
$$G({\bf r},{\bf r}')=\left(\frac{m}{ht}\right)^6\left[\overline{\left<n_{\bf Q}n_{\bf Q'} \right>}-\overline{\left<n_{\bf Q}\right>}\times\overline{\left<n_{\bf Q'}\right>}\right].$$
The disorder-averages of the momentum density distributions are easy to calculate. For instance, for the
two flavor superfluid state in Eq.~(\ref{eq:phistate}) we have
\begin{eqnarray}
\overline{\left<\Phi\right|n_{\bf Q}\left|\Phi\right>}&=&
\overline{|\tilde{\Psi}_x({\bf Q})|^2}+\overline{|\tilde{\Psi}_y({\bf Q})|^2}\nonumber\\
&+&2{\rm Re}\left[\overline{\tilde{\Psi}_x({\bf Q})^*\tilde{\Psi}_y({\bf Q})}\right].\label{eq:app4}
\end{eqnarray}
The averages in Eq.~({\ref{eq:app4}}) can be calculated using the representation in Eqs.~(\ref{eq:psirep1}) and (\ref{eq:psirep2})
\begin{eqnarray}
\overline{|\tilde{\Psi}_x|^2}
&=&
| \tilde{\phi}_0^x|^2{\frac{\rho}{2}}\sum_{\stackrel{m_1n_1}{m_2n_2}}e^{i{\bf Q}\cdot ({\bf R}_1-{\bf R}_2)}(-1)^{n_1+n_2}\overline{\alpha_{m_1n_1}\alpha_{m_2n_2}}.\nonumber
\end{eqnarray}
The average means averaging over all $\alpha_{mn}=\pm 1$ (and all $\beta_{mn}=\pm i$). It follows
$$\overline{|\tilde{\Psi}_x({\bf Q})|^2}={\frac{M}{2}}| \tilde{\phi}_0^x({\bf Q})|^2\quad \overline{|\tilde{\Psi}_x({\bf Q})|^2}={\frac{M}{2}}| \tilde{\phi}_0^y({\bf Q})|^2$$
and
$\overline{\tilde{\Psi}_x({\bf Q})^*\tilde{\Psi}_y({\bf Q})}=0.$ Hence, for the
two flavor case we find
\begin{equation}
\overline{\left<\Phi\right|n_{\bf Q}\left|\Phi\right>}={\frac{M}{2}}\left(| \tilde{\phi}_0^x({\bf Q})|^2
+|\tilde{\phi}_0^y({\bf Q})|^2\right),
\label{eq:app6}
\end{equation}
whereas for three flavors we have
$$\overline{\left<\Phi\right|n_{\bf Q}\left|\Phi\right>}={\frac{M}{3}}\left(| \tilde{\phi}_0^x({\bf Q})|^2
+|\tilde{\phi}_0^y({\bf Q})|^2+|\tilde{\phi}_0^z({\bf Q})|^2\right).$$

We now turn to the evaluation of the two point correlator which we begin by normal ordering
$$\overline{\left<n_{\bf Q}n_{\bf Q'}\right>}=(2\pi)^3\overline{\left<n_{\bf Q}\right>}\delta({\bf Q-Q'})+\overline{\left<\psi^\dagger_{\bf Q}\psi^\dagger_{\bf Q'}\psi_{\bf Q}\psi_{\bf Q'}\right>}.$$
The normal ordered expectation value $\left<\psi^\dagger_{\bf Q}\psi^\dagger_{\bf Q'}\psi_{\bf Q}\psi_{\bf Q'}\right>$ can be written in a form analogous to
Eq.~(\ref{eq:app1})
\begin{widetext}
\begin{eqnarray}
\left<\Phi\right|\psi_{\bf Q}^\dagger\psi_{\bf Q'}^\dagger\psi_{\bf Q}\psi_{\bf Q'}\left|\Phi\right>&=&\sum_{ijkl}
e^{i{\bf Q}\cdot({\bf R}_i-{\bf R}_j)}e^{i{\bf Q'}\cdot({\bf R}_k-{\bf R}_l)}\nonumber\\
&\times&\left<\Phi\Big|\left[X^\dagger_{i}(-1)^{n_i}{\tilde{\phi}_0^x}({\bf Q})^*+
Y^\dagger_{i}(-1)^{m_i}{\tilde{\phi}_0^y}({\bf Q})^*\right]\left[X^\dagger_{k}(-1)^{n_k}{\tilde{\phi}_0^x}({\bf Q'})^*+
Y^\dagger_{k}(-1)^{m_k}{\tilde{\phi}_0^y}({\bf Q'})^*\right]\right.\nonumber\\
&\times&\left.\left[X_{j}(-1)^{n_j}{\tilde{\phi}_0^x}({\bf Q})+
Y_{j}(-1)^{m_j}{\tilde{\phi}_0^y}({\bf Q})\right]\left[X_{l}(-1)^{n_l}{\tilde{\phi}_0^x}({\bf Q'})+
Y_{l}(-1)^{m_l}{\tilde{\phi}_0^y}({\bf Q'})\right]|\Phi\right>.\label{eq:app7}
\end{eqnarray}
\end{widetext}
Here the subscripts $ijkl$ are collective row and column coordinates for the site index in the 2D lattice.

For the two flavor
superfluid state in Eq.~(\ref{eq:phistate}) (a single plane with $N\times N$ sites having a total of $M$ particles) it is easy to verify that the expectation values of on-site operators are given by expressions of the type
$$\left<\Phi|X^\dagger_iY^\dagger_k Y_j X_l|\Phi\right>=\frac{M(M-1)}{4N^4}\alpha_i^*\beta_k^*\beta_j\alpha_l\approx \frac{\rho^2}{4}\alpha_i^*\beta_k^*\beta_j\alpha_l$$
To calculate the average over disorder we have to average over $\alpha_j=\pm1$ and $\beta_j=\pm i$. The nonzero averages are easily seen to be
\begin{eqnarray}
\overline{\left<X^\dagger_iX^\dagger_k X_j X_l\right>}&=&\frac{\rho^2}{4}\overline{\alpha_i^*\alpha_k^*\alpha_j\alpha_l}\label{eq:XXXX}\\
&=&\frac{\rho^2}{4}\left[\delta_{ik}\delta_{jl}+\delta_{ij}\delta_{kl}+\delta_{il}\delta_{kj}\right]\label{eq:XXYY}\\
\overline{\left<X^\dagger_iX^\dagger_k Y_j Y_l\right>}&=&\frac{\rho^2}{4}\overline{\alpha_i^*\alpha_k^*\beta_j\beta_l}
=-\frac{\rho^2}{4}\delta_{ik}\delta_{jl}\\
\overline{\left<Y^\dagger_iX^\dagger_k Y_j X_l\right>}&=&\frac{\rho^2}{4}\overline{\beta_i^*\alpha_k^*\beta_j\alpha_l}
=\frac{\rho^2}{4}\delta_{ij}\delta_{kl}\\
\overline{\left<Y^\dagger_iX^\dagger_k X_j Y_l\right>}&=&\frac{\rho^2}{4}\overline{\beta_i^*\alpha_k^*\alpha_j\beta_l}
=\frac{\rho^2}{4}\delta_{il}\delta_{kj}\\
\overline{\left<X^\dagger_iY^\dagger_k Y_j X_l\right>}&=&\frac{\rho^2}{4}\overline{\alpha_i^*\beta_k^*\beta_j\alpha_l}
=\frac{\rho^2}{4}\delta_{il}\delta_{kj}\\
\overline{\left<X^\dagger_iY^\dagger_k X_j Y_l\right>}&=&\frac{\rho^2}{4}\overline{\alpha_i^*\beta_k^*\alpha_j\beta_l}
=\frac{\rho^2}{4}\delta_{kl}\delta_{ij}\\
\overline{\left<Y^\dagger_iY^\dagger_k X_j X_l\right>}&=&\frac{\rho^2}{4}\overline{\beta_i^*\beta_k^*\alpha_j\alpha_l}
=-\frac{\rho^2}{4}\delta_{ik}\delta_{jl}\label{eq:YYXX}\\
\overline{\left<Y^\dagger_iY^\dagger_k Y_j Y_l\right>}&=&\frac{\rho^2}{4}\overline{\beta_i^*\beta_k^*\beta_j\beta_l}\nonumber\\
&=&\frac{\rho^2}{4}\left[\delta_{ik}\delta_{jl}+\delta_{ij}\delta_{kl}+\delta_{il}\delta_{kj}\right]\label{eq:YYYY}.
\end{eqnarray}
Note that Eqs.~(\ref{eq:XXXX}),(\ref{eq:XXYY}),(\ref{eq:YYXX}) and (\ref{eq:YYYY}) have contributions that
correspond to pairs of particles propagating. The disorder average of single particle propagation being zero due to the random orientation of phases. Using Eqs.~(\ref{eq:app7})-(\ref{eq:YYYY}) we can evaluate the terms
in Eq.~(\ref{eq:app7}) which are nonzero. We state each  term contributing to the correlator in Eq.~(\ref{eq:app7}) separately using subscripts to denote the specific
ordered combination of operators from which the term derives.
\begin{widetext}
\begin{eqnarray}
\overline{\left<\Phi\right|\psi_{\bf Q}^\dagger\psi_{\bf Q'}^\dagger\psi_{\bf Q}\psi_{\bf Q'}\left|\Phi\right>}_{XXXX}&=&|\phi_0^x({\bf Q})|^2|\phi_0^x({\bf Q}')|^2\frac{\rho^2}{4}\sum_{ij}\left[
1+e^{i({\bf Q}+{\bf Q}')\cdot({\bf R}_i-{\bf R}_j)}+e^{i({\bf Q}-{\bf Q}')\cdot({\bf R}_i-{\bf R}_j)}\right]\\
\overline{\left<\Phi\right|\psi_{\bf Q}^\dagger\psi_{\bf Q'}^\dagger\psi_{\bf Q}\psi_{\bf Q'}\left|\Phi\right>}_{XXYY}&=&-{\tilde{\phi}_0^x}({\bf Q})^*{\tilde{\phi}_0^x}({\bf Q'})^*{\tilde{\phi}_0^y}({\bf Q}){\tilde{\phi}_0^y}({\bf Q'})\frac{\rho^2}{4}\sum_{ij}
e^{i({\bf Q}+{\bf Q}')\cdot({\bf R}_i-{\bf R}_j)}\\
\overline{\left<\Phi\right|\psi_{\bf Q}^\dagger\psi_{\bf Q'}^\dagger\psi_{\bf Q}\psi_{\bf Q'}\left|\Phi\right>}_{YXYX}&=&{\tilde{\phi}_0^y}({\bf Q})^*{\tilde{\phi}_0^x}({\bf Q'})^*{\tilde{\phi}_0^y}({\bf Q}){\tilde{\phi}_0^x}({\bf Q'})\frac{M^2}{4}\\
\overline{\left<\Phi\right|\psi_{\bf Q}^\dagger\psi_{\bf Q'}^\dagger\psi_{\bf Q}\psi_{\bf Q'}\left|\Phi\right>}_{YXXY}&=&{\tilde{\phi}_0^y}({\bf Q})^*{\tilde{\phi}_0^x}({\bf Q'})^*{\tilde{\phi}_0^x}({\bf Q}){\tilde{\phi}_0^y}({\bf Q'})\frac{\rho^2}{4}\sum_{ij}
e^{i({\bf Q}-{\bf Q}')\cdot({\bf R}_i-{\bf R}_j)}\\
\overline{\left<\Phi\right|\psi_{\bf Q}^\dagger\psi_{\bf Q'}^\dagger\psi_{\bf Q}\psi_{\bf Q'}\left|\Phi\right>}_{XYYX}&=&{\tilde{\phi}_0^x}({\bf Q})^*{\tilde{\phi}_0^y}({\bf Q'})^*{\tilde{\phi}_0^y}({\bf Q}){\tilde{\phi}_0^x}({\bf Q'})\frac{\rho^2}{4}\sum_{ij}
e^{i({\bf Q}-{\bf Q}')\cdot({\bf R}_i-{\bf R}_j)}\\
\overline{\left<\Phi\right|\psi_{\bf Q}^\dagger\psi_{\bf Q'}^\dagger\psi_{\bf Q}\psi_{\bf Q'}\left|\Phi\right>}_{XYXY}&=&{\tilde{\phi}_0^x}({\bf Q})^*{\tilde{\phi}_0^y}({\bf Q'})^*{\tilde{\phi}_0^x}({\bf Q}){\tilde{\phi}_0^y}({\bf Q'})\frac{M^2}{4}\\
\overline{\left<\Phi\right|\psi_{\bf Q}^\dagger\psi_{\bf Q'}^\dagger\psi_{\bf Q}\psi_{\bf Q'}\left|\Phi\right>}_{YYXX}&=&-{\tilde{\phi}_0^y}({\bf Q})^*{\tilde{\phi}_0^y}({\bf Q'})^*{\tilde{\phi}_0^x}({\bf Q}){\tilde{\phi}_0^x}({\bf Q'})\frac{\rho^2}{4}\sum_{ij}
e^{i({\bf Q}+{\bf Q}')\cdot({\bf R}_i-{\bf R}_j)}\\
\overline{\left<\Phi\right|\psi_{\bf Q}^\dagger\psi_{\bf Q'}^\dagger\psi_{\bf Q}\psi_{\bf Q'}\left|\Phi\right>}_{YYYY}&=&|\phi_0^y({\bf Q})|^2|\phi_0^y({\bf Q}')|^2\frac{\rho^2}{4}\sum_{ij}\left[
1+e^{i({\bf Q}+{\bf Q}')\cdot({\bf R}_i-{\bf R}_j)}+e^{i({\bf Q}-{\bf Q}')\cdot({\bf R}_i-{\bf R}_j)}\right].\label{eq:app8}
\end{eqnarray}
Collecting the results of Eqs.~(\ref{eq:app6})-(\ref{eq:app8}) we find
\begin{eqnarray}
G_{SF}^{2D}({\bf r},{\bf r}')&\propto&(2\pi)^3 \frac{M}{2}\delta({\bf Q-Q'})\left(| \tilde{\phi}_0^x({\bf Q})|^2
+|\tilde{\phi}_0^y({\bf Q})|^2\right)
+\frac{\rho^2}{4}\left|\tilde{\phi_0^x}({\bf Q})\tilde{\phi_0^x}({\bf Q'})+\tilde{\phi_0^y}({\bf Q})\tilde{\phi_0^y}({\bf Q'})\right|^2\sum_{ij}e^{i({\bf Q}-{\bf Q}')\cdot({\bf R}_i-{\bf R}_j)}\nonumber\\
&+&\frac{\rho^2}{4}
\left|\tilde{\phi_0^x}({\bf Q})\tilde{\phi_0^x}({\bf Q'})-\tilde{\phi_0^y}({\bf Q})\tilde{\phi_0^y}({\bf Q'})\right|^2
\sum_{ij}e^{i({\bf Q}+{\bf Q}')\cdot({\bf R}_i-{\bf R}_j)}\label{eq:GSF2D}
\end{eqnarray}
\end{widetext}
where the factor of proportionality is $\left(\frac{m}{ht}\right)^6$. The Fourier sums give, in the limit of an infinite lattice, sequences of delta functions
\begin{eqnarray}
\sum_{ij}e^{i({\bf Q}\pm{\bf Q}')\cdot({\bf R}_i-{\bf R}_j)}\rightarrow\left(\frac{2\pi N}{a}\right)^2 \sum_{i}\delta([{\bf Q\pm Q'}]-{\bf G}_i)\nonumber
\end{eqnarray}
where ${\bf G}_i$ are reciprocal lattice vectors. The most interesting part of Eq.~(\ref{eq:GSF2D}) is the second line
which comes from the pair like propagation. This can be used as a signature to detect the superfluid phase even if thermal disorder has restored the $Z_2$ symmetry.

For comparison we also look at the 2D
(two flavors)
Mott state. For simplicity we consider unit filling. There are three scenarios
for the unit filling Mott state that need to be considered; (a) FM Mott
state, \emph{i.e.} all atoms of the same flavor (b) AFM Mott state,
X-flavor and Y-flavor on alternating sites and (c) thermally
disordered Mott state with random occupation of X- and Y- flavor on
each site.

In the ferromagnetic Mott state at unit filling $M=N^2$ we have two degenerate ground states $\left|\Phi_1\right>=\prod_{i}X^\dagger_i\left|0\right>$
and $\left|\Phi_2\right>=\prod_{i}Y^\dagger_i\left|0\right>$
and the average in Eq.~(\ref{eq:disorderaverage}) is trivial to evaluate
\begin{eqnarray}
\overline{\left<\Phi\right|n_{\bf Q}\left|\Phi\right>}&=&\frac{1}{2}\left(\left<\Phi_1\right|n_{\bf Q}\left|\Phi_1\right>+\left<\Phi_2\right|n_{\bf Q}\left|\Phi_2\right>\right)\nonumber\\
&=&\frac{M}{2}\left(|\phi_0^x({\bf Q})|^2+|\phi_0^y({\bf Q})|^2\right).
\end{eqnarray}
The momentum correlator can be calculated using Eq.~(\ref{eq:app7}). For the state $\left|\Phi_1\right>$ this equation reduces to
\begin{widetext}
\begin{eqnarray}
\left<\Phi_1\right|\psi_{\bf Q}^\dagger\psi_{\bf Q'}^\dagger\psi_{\bf Q}\psi_{\bf Q'}\left|\Phi_1\right>&=&|{\tilde{\phi}_0^x}({\bf Q})|^2|{\tilde{\phi}_0^x}({\bf Q'})|^2\sum_{ijkl}
e^{i{\bf Q}\cdot({\bf R}_i-{\bf R}_j)}e^{i{\bf Q'}\cdot({\bf R}_k-{\bf R}_l)}\left<\Phi_1\right|X^\dagger_{i}X^\dagger_{k}X_{j}X_{l}\left|\Phi_1\right>.\label{eq:app10}
\end{eqnarray}
There are two pairings of operators that contribute to the average
$$\left<\Phi_1\right|X^\dagger_{i}X^\dagger_{k}X_{j}X_{l}\left|\Phi_1\right>=(1-\delta_{ik})(\delta_{ij}\delta_{kl}+\delta_{il}\delta_{kj}).$$
The term $(1-\delta_{ik})$ results from having only one particle at each site but we will ignore this term (and terms similar to it in what follows)
since its relative contribution is of order $1/N^2$. The disorder average contains only two states yielding
\begin{eqnarray}
\overline{\left<\Phi\right|\psi_{\bf Q}^\dagger\psi_{\bf
Q'}^\dagger\psi_{\bf Q}\psi_{\bf
Q'}\left|\Phi\right>}&=&\frac{1}{2}\left(|{\tilde{\phi}_0^x}({\bf
Q})|^2|{\tilde{\phi}_0^x}({\bf Q'})|^2+|{\tilde{\phi}_0^y}({\bf
Q})|^2|{\tilde{\phi}_0^y}({\bf Q'})|^2\right)\sum_{ik}\left[
1+e^{i({\bf Q-Q'})\cdot({\bf R}_i-{\bf
R}_k)}\right].\label{eq:app10b}
\end{eqnarray}
The correlation function for the Ferromagnetic Mott state is thus
\begin{eqnarray}
G_{FM}^{2D}({\bf r},{\bf r'})&\propto&(2\pi)^3\delta({\bf Q-Q'})\frac{M}{2}\left(|\phi_0^x({\bf Q})|^2+|\phi_0^y({\bf Q})|^2\right)+\frac{M^2}{4}\left(|\phi_0^x({\bf Q})|^2-|\phi_0^y({\bf Q})|^2\right)\left(|\phi_0^x({\bf Q'})|^2-|\phi_0^y({\bf Q'})|^2\right)\nonumber\\
&+&\frac{1}{2}\left(|{\tilde{\phi}_0^x}({\bf Q})|^2|{\tilde{\phi}_0^x}({\bf Q'})|^2+|{\tilde{\phi}_0^y}({\bf Q})|^2|{\tilde{\phi}_0^y}({\bf Q'})|^2\right)\sum_{ik}
e^{i({\bf Q-Q'})\cdot({\bf R}_i-{\bf R}_k)}.\label{eq:GFM2D}
\end{eqnarray}
\end{widetext}
In the antiferromagetic Mott state the disorder average is again over two states. Dividing the 2D lattice into two sublattices $A$ and $B$ these states are $\left|\Phi_1\right>=\prod_{i\in A}\prod_{j\in B}X^\dagger_iY^\dagger_j\left|0\right>$ and $\left|\Phi_2\right>=\prod_{j\in A}\prod_{i\in B}X^\dagger_iY^\dagger_j\left|0\right>$.
For the momentum density we have
\begin{eqnarray}
\left<\Phi_1|n_{\bf Q}|\Phi_1\right>&=&\left<\Phi_2|n_{\bf Q}|\Phi_2\right>=\frac{M}{2}\left[|{\tilde{\phi}_0^x}({\bf Q})|^2+
|{\tilde{\phi}_0^y}({\bf Q})|^2\right]\nonumber
\end{eqnarray}
hence we find again that
$$\overline{\left<\Phi|n_{\bf Q}|\Phi\right>}=\frac{M}{2}\left[|{\tilde{\phi}_0^x}({\bf Q})|^2+
|{\tilde{\phi}_0^y}({\bf Q})|^2\right].$$
The normal ordered two point correlator can again be written in the form of Eq.~(\ref{eq:app7})
and has 6 nonzero contributions. The disorder average over the two states will in this case make no
difference since the two different states always give the same contribution and it is enough to consider one of them.
\begin{widetext}
\begin{eqnarray}
\left<\Phi_1\right|\psi_{\bf Q}^\dagger\psi_{\bf Q'}^\dagger\psi_{\bf Q}\psi_{\bf Q'}\left|\Phi_1\right>_{XXXX}&=&|{\tilde{\phi}_0^x}({\bf Q})|^2|{\tilde{\phi}_0^x}({\bf Q'})|^2\left(\frac{M^2}{4}+\sum_{i\in Aj\in A}
e^{i({\bf Q-Q'})\cdot({\bf R}_i-{\bf R}_j)}\right)\\
\left<\Phi_1\right|\psi_{\bf Q}^\dagger\psi_{\bf Q'}^\dagger\psi_{\bf Q}\psi_{\bf Q'}\left|\Phi_1\right>_{YYYY}&=&
|{\tilde{\phi}_0^y}({\bf Q})|^2|{\tilde{\phi}_0^y}({\bf Q'})|^2\left(\frac{M^2}{4}+\sum_{i\in Bj\in B}
e^{i({\bf Q-Q'})\cdot({\bf R}_i-{\bf R}_j)}\right)\\
\left<\Phi_1\right|\psi_{\bf Q}^\dagger\psi_{\bf Q'}^\dagger\psi_{\bf Q}\psi_{\bf Q'}\left|\Phi_1\right>_{XYXY}&=&\frac{M^2}{4}
|{\tilde{\phi}_0^x}({\bf Q})|^2
|{\tilde{\phi}_0^y}({\bf Q'})|^2\\
\left<\Phi_1\right|\psi_{\bf Q}^\dagger\psi_{\bf Q'}^\dagger\psi_{\bf Q}\psi_{\bf Q'}\left|\Phi_1\right>_{XYYX}&=&{\tilde{\phi}_0^x}({\bf Q})^*
{\tilde{\phi}_0^y}({\bf Q'})^*
{\tilde{\phi}_0^y}({\bf Q}){\tilde{\phi}_0^x}({\bf Q'})\sum_{i\in Aj\in B}
e^{i({\bf Q-Q'})\cdot({\bf R}_i-{\bf R}_j)}\\
\left<\Phi_1\right|\psi_{\bf Q}^\dagger\psi_{\bf Q'}^\dagger\psi_{\bf Q}\psi_{\bf Q'}\left|\Phi_1\right>_{YXYX}&=&\frac{M^2}{4}
|{\tilde{\phi}_0^y}({\bf Q})|^2|{\tilde{\phi}_0^x}({\bf Q'})|^2\\
\left<\Phi_1\right|\psi_{\bf Q}^\dagger\psi_{\bf Q'}^\dagger\psi_{\bf Q}\psi_{\bf Q'}\left|\Phi_1\right>_{YXXY}&=&{\tilde{\phi}_0^y}({\bf Q})^*{\tilde{\phi}_0^x}({\bf Q'})^*{\tilde{\phi}_0^x}({\bf Q})
{\tilde{\phi}_0^y}({\bf Q'})\sum_{i\in A j\in B }
e^{i({\bf Q-Q')}\cdot({\bf R}_j-{\bf R}_i)}
\end{eqnarray}
Hence, we find for the 2D antiferromagnetic Mott state at unit filling the correlation function
\begin{eqnarray}
G_{AFM}^{2D}({\bf r},{\bf r'})&\propto&(2\pi)^3\delta({\bf Q-Q'})\frac{M}{2}\left[|{\tilde{\phi}_0^x}({\bf Q})|^2+
|{\tilde{\phi}_0^y}({\bf Q})|^2\right]\nonumber\\
&+&2{\rm Re}\left[{\tilde{\phi}_0^x}({\bf Q})^*
{\tilde{\phi}_0^y}({\bf Q'})^*
{\tilde{\phi}_0^y}({\bf Q}){\tilde{\phi}_0^x}({\bf Q'})\right]\sum_{i\in Aj\in B}
\cos\left[{({\bf Q-Q'})\cdot({\bf R}_i-{\bf R}_j)}\right]\nonumber\\
&+&\left(|{\tilde{\phi}_0^x}({\bf Q})|^2|{\tilde{\phi}_0^x}({\bf Q'})|^2+|{\tilde{\phi}_0^y}({\bf Q})|^2|{\tilde{\phi}_0^y}({\bf Q'})|^2\right)\sum_{i\in Aj\in A}
e^{i({\bf Q-Q'})\cdot({\bf R}_i-{\bf R}_j)}.\label{eq:GAFM2D}
\end{eqnarray}
The Fourier sums converges in the limit of large $N$ to
$$\sum_{i\in Aj\in B}
\cos\left[{({\bf Q-Q'})\cdot({\bf R}_i-{\bf R}_j)}\right]=\frac{\pi^2N^2}{2a^2}\sum_{mn}[(-1)^n+(-1)^m]\delta(Q_x-\frac{n\pi}{a})\delta(Q_y-\frac{n\pi}{a})$$
$$\sum_{i\in Aj\in A}
e^{i({\bf Q-Q'})\cdot({\bf R}_i-{\bf R}_j)}=\frac{\pi^2N^2}{2a^2}\sum_{mn}[1+(-1)^{n+m}]\delta(Q_x-\frac{n\pi}{a})\delta(Q_y-\frac{n\pi}{a})$$

and the correlation function for the antiferromagnetic Mott state will thus have peaks at locations corresponding to half reciprocal lattice vectors.
\end{widetext}
Finally we look at the disordered Mott state where each site holds one atom but whether it is an $X$ or a $Y$ is random. The manifold of states to average over thus contains ${\cal N}=2^{N^2}$ states. Such a state can be written as
$$\left|\Phi\right>=\prod_i\frac{1}{2}\left[X^\dagger_i(1+\eta_i)+Y^\dagger_i(1-\eta_i)\right]\left|0\right>$$
where $\eta_i$ is a random field taking on values $\pm 1$ on each site $i$.
The disorder averaged momentum distribution is again the same as before
\begin{eqnarray}
\overline{\left<\Phi\right|\psi_{\bf Q}^\dagger\psi_{\bf Q}\left|\Phi\right>}&=&\sum_{i}\frac{1+\overline{\eta_i}}{2}{\tilde{\phi}_0^x}|({\bf Q})|^2+
\frac{1-\overline{\eta_i}}{2}|{\tilde{\phi}_0^y}({\bf Q})|^2\nonumber\\
&=&\frac{M}{2}\left({\tilde{\phi}_0^x}|({\bf Q})|^2+
|{\tilde{\phi}_0^y}({\bf Q})|^2\right)
\end{eqnarray}
and there are six contributions to momentum correlator
\begin{widetext}
\begin{eqnarray}
\overline{\left<\Phi\right|\psi_{\bf Q}^\dagger\psi_{\bf Q'}^\dagger\psi_{\bf Q}\psi_{\bf Q'}\left|\Phi\right>}_{XXXX}&=&\frac{1}{4}|{\tilde{\phi}_0^x}({\bf Q})|^2|{\tilde{\phi}_0^x}({\bf Q'})|^2\sum_{ik}1+
e^{i({\bf Q-Q'})\cdot({\bf R}_i-{\bf R}_k)}\\
\overline{\left<\Phi\right|\psi_{\bf Q}^\dagger\psi_{\bf Q'}^\dagger\psi_{\bf Q}\psi_{\bf Q'}\left|\Phi\right>}_{YYYY}&=&\frac{1}{4}|{\tilde{\phi}_0^y}({\bf Q})|^2|{\tilde{\phi}_0^y}({\bf Q'})|^2\sum_{ik}1+
e^{i({\bf Q-Q'})\cdot({\bf R}_i-{\bf R}_k)}\\
\overline{\left<\Phi\right|\psi_{\bf Q}^\dagger\psi_{\bf Q'}^\dagger\psi_{\bf Q}\psi_{\bf Q'}\left|\Phi\right>}_{XYXY}&=&\frac{M^2}{4}|{\tilde{\phi}_0^x}({\bf Q})|^2
|{\tilde{\phi}_0^y}({\bf Q'})|^2\\
\overline{\left<\Phi\right|\psi_{\bf Q}^\dagger\psi_{\bf Q'}^\dagger\psi_{\bf Q}\psi_{\bf Q'}\left|\Phi\right>}_{XYYX}&=&\frac{1}{4}{\tilde{\phi}_0^x}({\bf Q})^*
{\tilde{\phi}_0^y}({\bf Q'})^*
{\tilde{\phi}_0^y}({\bf Q}){\tilde{\phi}_0^x}({\bf Q'})\sum_{ik}
e^{i({\bf Q-Q'})\cdot({\bf R}_i-{\bf R}_k)}\\
\overline{\left<\Phi\right|\psi_{\bf Q}^\dagger\psi_{\bf Q'}^\dagger\psi_{\bf Q}\psi_{\bf Q'}\left|\Phi\right>}_{YXYX}&=&\frac{M^2}{4}
|{\tilde{\phi}_0^y}({\bf Q})|^2|{\tilde{\phi}_0^x}({\bf Q'})|^2
\\
\overline{\left<\Phi\right|\psi_{\bf Q}^\dagger\psi_{\bf Q'}^\dagger\psi_{\bf Q}\psi_{\bf Q'}\left|\Phi\right>}_{YXXY}&=&\frac{1}{4}{\tilde{\phi}_0^y}({\bf Q})^*{\tilde{\phi}_0^x}({\bf Q'})^*{\tilde{\phi}_0^x}({\bf Q})
{\tilde{\phi}_0^y}({\bf Q'})\sum_{ik}
e^{i({\bf Q-Q')}\cdot({\bf R}_i-{\bf R}_k)}
\end{eqnarray}
resulting in a correlation function for the disordered Mott state
\begin{eqnarray}
G_{DO}^{2D}({\bf r},{\bf r'})&\propto&(2\pi)^3\delta({\bf Q-Q'})\frac{M}{2}\left[|{\tilde{\phi}_0^x}({\bf Q})|^2+
|{\tilde{\phi}_0^y}({\bf Q})|^2\right]+\frac{1}{4}|{\tilde{\phi}_0^x}({\bf Q})^*{\tilde{\phi}_0^x}({\bf Q'})+{\tilde{\phi}_0^y}({\bf Q})^*{\tilde{\phi}_0^y}({\bf Q'})|^2\sum_{ik}
e^{i({\bf Q-Q'})\cdot({\bf R}_i-{\bf R}_k)}.\nonumber\\
\label{eq:GDO2D}
\end{eqnarray}
\end{widetext}
\subsection{Correlations 3D, three flavors, $T\gtrsim 0$} We now look at the
momentum correlations in the thermally disordered
three flavor superfluid
phase. To evaluate the correlation function
$\overline{\left<\Phi\right|\psi_{\bf Q}^\dagger\psi_{\bf
Q'}^\dagger\psi_{\bf Q}\psi_{\bf Q'}\left|\Phi\right>}$ one can
write down the extension of Eq.~(\ref{eq:app7}). There will be
overall 81 terms in the expansion to evaluate. When taking the
disorder average only 21 terms are nonzero. Note that when taking
the disorder average in
the three flavor model
one has to average not only over all
possible $\pi$ flips of the phases but also over the symmetry
breaking field $\sigma_m$ (see Eq.~(\ref{eq:chiralfield})) to
account for the chiral symmetry breaking as well as over the 3
directions directions in which chiral symmetry is broken. The
nonzero averages one obtains in this way are shown below
\begin{eqnarray}
\overline{\left<X_i^\dagger X^\dagger_k X_j X_l\right>}&=&\overline{\left<Y_i^\dagger Y^\dagger_k Y_j Y_l\right>}=\overline{\left<Z_i^\dagger Z^\dagger_k Z_j Z_l\right>}\nonumber\\
&=&\frac{\rho^2}{9}[\frac{1}{2}\delta_{ik}\delta_{jl}+\delta_{ij}\delta_{kl}+\delta_{il}\delta_{kj}]\\
\overline{\left<X_i^\dagger X^\dagger_k Y_j Y_l\right>}&=&
\overline{\left<X_i^\dagger X^\dagger_k Z_j Z_l\right>}=-\frac{1}{4}\frac{\rho^2}{9}\delta_{ik}\delta_{jl}\\
\overline{\left<Y_i^\dagger Y^\dagger_k X_j X_l\right>}&=&
\overline{\left<Y_i^\dagger Y^\dagger_k Z_j Z_l\right>}=-\frac{1}{4}\frac{\rho^2}{9}\delta_{ik}\delta_{jl}\\
\overline{\left<Z_i^\dagger Z^\dagger_k Y_j Y_l\right>}&=&
\overline{\left<Z_i^\dagger Z^\dagger_k X_j X_l\right>}=-\frac{1}{4}\frac{\rho^2}{9}\delta_{ik}\delta_{jl}\\
\overline{\left<X_i^\dagger Y^\dagger_k X_j Y_l\right>}&=&
\overline{\left<X_i^\dagger Z^\dagger_k X_j Z_l\right>}=\frac{\rho^2}{9}\delta_{ij}\delta_{kl}\\
\overline{\left<Y_i^\dagger X^\dagger_k Y_j X_l\right>}&=&
\overline{\left<Y_i^\dagger Z^\dagger_k Y_j Z_l\right>}=\frac{\rho^2}{9}\delta_{ij}\delta_{kl}\\
\overline{\left<Z_i^\dagger X^\dagger_k Z_j X_l\right>}&=&
\overline{\left<Z_i^\dagger Y^\dagger_k Z_j Y_l\right>}=\frac{\rho^2}{9}\delta_{ij}\delta_{kl}\\
\overline{\left<X_i^\dagger Y^\dagger_k Y_j X_l\right>}&=&
\overline{\left<X_i^\dagger Z^\dagger_k Z_j X_l\right>}=\frac{\rho^2}{9}\delta_{il}\delta_{kj}\\
\overline{\left<Y_i^\dagger X^\dagger_k X_j Y_l\right>}&=&
\overline{\left<Y_i^\dagger Z^\dagger_k Z_j Y_l\right>}=\frac{\rho^2}{9}\delta_{il}\delta_{kj}\\
\overline{\left<Z_i^\dagger X^\dagger_k X_j Z_l\right>}&=&
\overline{\left<Z_i^\dagger Y^\dagger_k Y_j Z_l\right>}=\frac{\rho^2}{9}\delta_{il}\delta_{kj}
\end{eqnarray}
and the desired correlator is obtained
\begin{widetext}
\begin{eqnarray}
G_{SF}^{3D}({\bf r},{\bf r'})&\propto&(2\pi)^3\delta({\bf Q-Q'})\frac{M}{3}\left[|{\tilde{\phi}_0^x}({\bf Q})|^2+
|{\tilde{\phi}_0^y}({\bf Q})|^2+|{\tilde{\phi}_0^z}({\bf Q})|^2\right]+\frac{1}{2}\frac{\rho^2}{9}\left(|\tilde{\phi}_0^x({\bf Q})\tilde{\phi}_0^x({\bf Q'})-\tilde{\phi}_0^y({\bf Q})\tilde{\phi}_0^y({\bf Q'})|^2\right.\nonumber\\
&+&\left.|\tilde{\phi}_0^y({\bf Q})\tilde{\phi}_0^y({\bf Q'})-\tilde{\phi}_0^z({\bf Q})\tilde{\phi}_0^z({\bf Q'})|^2+|\tilde{\phi}_0^z({\bf Q})\tilde{\phi}_0^z({\bf Q'})-\tilde{\phi}_0^x({\bf Q})\tilde{\phi}_0^x({\bf Q'})|^2\right)\sum_{ik}
e^{i({\bf Q+Q'})\cdot({\bf R}_i-{\bf R}_k)}\nonumber\\
&+&\frac{\rho^2}{9}|\tilde{\phi}_0^x({\bf Q})\tilde{\phi}_0^x({\bf Q'})^*+\tilde{\phi}_0^y({\bf Q})\tilde{\phi}_0^y({\bf Q'})^*+\tilde{\phi}_0^z({\bf Q})\tilde{\phi}_0^z({\bf Q'})^*|^2\sum_{ik}e^{i({\bf Q-Q'})\cdot({\bf R}_i-{\bf R}_k)}.
\label{eq:GDO3D}
\end{eqnarray}
\end{widetext}

\end{document}